\documentclass[a4paper,11pt]{article}
\usepackage{jcappub} % for details on the use of the package, please
\usepackage[T1]{fontenc} % if needed
\usepackage{graphicx}
\usepackage{float}
\usepackage[margin=1in]{geometry}
\usepackage{color}

\def\1{\mbox{I\hspace{-.15em}1}}

\def\b{\begin{equation}}
\def\e{\end{equation}}
\def\bee{\begin{enumerate}}
\def\eee{\end{enumerate}}
\usepackage{titlesec}
\usepackage[titletoc,toc,title]{appendix} %appendix

\titleformat{\section}{\bfseries}{\thesection.}{0.5em}{} %appendix
\titleformat{\subsection}{\normalfont\itshape}{\thesubsection.}{0.5em}{} %appendix
\titleformat{\subsubsection}{\normalfont\itshape}{\thesubsubsection.}{0.6em}{} %appendix

\title{\boldmath Are Cold Dynamical Dark Energy Models Distinguishable in the Light of the Data? }

\author[a,1]{Aghileh S. Ebrahimi \note{aghileh@grad.kashanu.ac.ir}}
\author[a]{M. Monemzadeh}
\author[b]{H. Moshafi}

\affiliation[a]{Department of Physics of University of Kashan, Ravand, Kashan,  Iran}
\affiliation[b] {Ibn-Sina multidisciplinary Lab., Department of Physics, Shahid Beheshti University,
Velenjak, Tehran 19839, Iran}

\abstract{In this paper we obtain observational constraints on three dynamical cold dark energy models ,include PL , CPL and FSL, with most recent cosmological data and investigate their implication for structure formation, dark energy clustering and abundance of CMB local peaks. From  the joint analysis of the CMB temperature  power spectrum from observation of the {\it Planck}, SNIa light-curve, baryon acoustic oscillation, $f\sigma_8$ for large scale structure observations and the Hubble parameter, we find that $\Omega_{\rm
DE}=0.6862\pm0.0078$, $\alpha=0.1013\pm0.0031$ and $w_0=-1.3799^{+0.0036}_{-0.0028}$ for the PL model, $\Omega_{\rm
DE}=0.6880^{+0.0100}_{-0.0079}$, $w_0=-1.08045^{+0.00041}_{-0.00062}$  and $w_1=-0.12190^{+0.00050}_{-0.00030}$ for the CPL model and $\Omega_{\rm DE}=0.6893\pm0.0078$, $w_0=-0.9994\pm0.0076$ and $w_1=-0.0082^{+0.0044}_{-0.0051}$ for the FSL model at $1\sigma$ confidence interval. The PL model has  matter-like contribution to the energy content of early universe due to crossing behavior of its Equation of state. Therefore, the PL model has the highest growth of matter density, $ \Delta_{m} $, and matter power spectrum, $ P(k) $, compared to $\Lambda$CDM and other models. For the CPL on the other hand, the structure formation is considerably suppressed while the FSL has behavior similar to standard model of cosmology. Studying the clustering of dark energy, $ \Delta_{DE}$, yields positive but small value with maximum of $ \Delta_{DE}\simeq10^{-3} $ at early time due to matter behaviour of the PL, while for the CPL and FSL cross $ \Delta_{DE}=0 $ several time which demonstrate  void of dark energy with $ \Delta_{DE}\simeq-10^{-11} $ in certain periods of the history of dark energy evolution. Among these three models, the PL model demonstrate that is more compatible with $ f\sigma_{8} $ data. We also investigated a certain geometrical measure, namely the abundance of local maxima as a function of threshold for three DDE models and find that the method is potentially capable to discriminate between the models, especially far from mean threshold. The contribution of PL and CPL for late ISW are significant compared to cosmological constant and FSL model. The tension in the Hubble parameters is almost alleviated in the PL model. }

\begin{document}
\maketitle
\flushbottom

\section{Introduction}
\label{sec:intro}

The accelerating expansion of the universe was first noticed by  Riess et al. \cite{Riess:1998cb} in High-redshift Supernova Search Team and by Perlumuter et al. \cite{Perlmutter:1998np} in Supernova Cosmology Project Team, independently. After that, many scientific projects have been established in order to asses the source of this phenomenon as well as to  achieve  desired observational accuracy. Several observational achievements including Cosmic Microwave Background Anisotropy (CMB), Large Scale Structure (LSS), Baryonic Acoustic Oscillations (BAO) and indirect estimations for the Hubble parameter versus redshift also strongly support mentioned dynamics of our cosmos on large scales with high precision \cite{BAO-H-1, BAO-H-2}.
In spite of considerable researchers focused on the nature of this accelerated expansion, there are many challenges in clarifying the corresponding source(s). The historically well known possibility is cosmological constant (CC), so-called concordance model.  Although  being a greatly successful scenario with high value of confidence interval, $\Lambda$CDM model suffer some problems are still pending to resolve.
Among them are the  sharp transition from the cold dark matter era to $\Lambda$ dominant epoch and fine tuning problems \cite{CCP,Kunz:2012aw}.  In addition, has also been recent report on flowing tension: the excessive congestion of matter and the missing satellites puzzle \cite{moore99,Klypin99}, the cusp-core problem \cite{Navarro96,Blok10,Donato09} and in a reliable observational projects run by the $Planck$ collaboration, the observed clusters are fewer than expected \cite{Adecluster16}.

From a robust point of view,  there are three approaches beyond
the$\Lambda$CDM model to elucidate accelerating epoch: The first
approach are theoretical orientation and phenomenological dark
fluids resulting in dynamical dark energy (DDE) also known as
quintessence \cite{Kamenshchik:2001cp}, non canonical scalar field
(k-essence) \cite{ArmendarizPicon:2000ah}, phantom
\cite{Caldwell:1999ew}, coupled dark energy \cite{Amendola:1999er}
and so on. The second approach devoted to modification of gravity
such as $f(R)$ gravity \cite{Carroll:2003wy,Nojiri:2003ft},
scalar-tensor theories \cite{Amendola:1999qq,Uzan:1999ch}. The third
approach, dedicated to thermodynamics point of view to propose
phenomenological exotic fluids
\citep{Copeland:2006wr,Amendola:2012ys,Horndeski:1974wa,Konnig:2016idp,Bento:2002ps,Mostaghel:2016lcd,Nojiri:2010wj,Bull:2015stt}.
Although, a more recent theory of Uber Gravity devoted to ensemble
average on all consistent models of gravity \cite{nima16,nima17}. In
this work we focus on DDE models.

Reconstruction of equation of state is another viewpoint based on observations including parametric and non-parametric approaches. Parametric reconstruction is based upon estimation of model parameters from different observational data sets \cite{Huterer:1998qv}, while in non-parametric reconstruction , find nature of cosmic evolution from observation rather than any prior assumptions of parametric from any cosmological parameters. \cite{Sahlen:2005zw,Nair:2013sna,Holsclaw}.

Since  CC equation of state is  precisely $ w=-1 $ and recent data
slightly favorite $ w<-1 $, small deviation from CC allow one to
consider models with $ w\neq-1 $.  According to constant nature of
CC, it is contributed directly on background evolution. Any
conceivable assessment of DDE model needs to incorporate not only
the characterizing background evolution but also various aspect of
perturbations leading to structure formation affected by DDE. The
robust perturbations analysis may provide opportunity to
discriminate various scenarios for dark energy. It is worth noting
that, the nature of dark energy can be investigated not only by
corresponding equation of state but also through sound speed and
associated clustering. The measurement of $3D$ gravitational
perturbation of galaxy redshift survey at enough large scale
suggested for probing of dark energy clustering.
 Dark energy both clustering individually and cluster together with matter at large spacial scale \cite{DEC-4}. At late time, it is found that clustering (overdensity) of matter correspond to void (underdensity) of dark energy, conversely a void in matter was seen produce a local DDE overdensity \cite{DEC-9}. Furthermore, for $ w<-1 $ dark energy void expected and $ w>-1 $ dark energy cluster happened \cite{DEC-5} howbeit there are analysis which $ \Delta_{DE} >0 $ occur for $ w>-1 $ \cite{DEC-9,31DEC-5}.
The signature of cold dark energy on galaxy cluster abundance explored \cite{DEC-1}.

A key question in dealing with model of dark energy family, is finding the realistic scenario. Practically, the proposed models to explain the accelerating expansion of the universe are not distinguishable from $\Lambda$CDM at the level of background expansion history, but hopefully, considering structure formation even at linear regime would lead to different consequences. Currently, precision limitation on current observational data prevent to achieve discrimination between the models.  Meanwhile, there are many attempts dedicated to mentioned goal that
use most recent high resolution observations in one hand, and introducing curious criteria ranging from new observables to robust topological and geometrical measures  in another hand \cite{jain03,taylor07,Kitching08}. In methods such as State-finder \cite{Sahni:2002fz} and {\it Om} diagnostic \cite{Arabsalmani:2011fz}, the corresponding parameters are not observable quantities.
Part of current data such as ISW and the matter power spectrum do not have enough precision to rule out models with behaviors similar to $\Lambda$CDM. The forthcoming observations designed to constrain the growth function are very promising\cite{FutureGrowsFuvnction}.
 Future galaxy weak lensing experiments such as the Dark Energy Survey (DES) \cite{DES} and Euclid \cite{Euclid} and SKA project \cite{SKA} will  potentially be able to discriminate between the $\Lambda$CDM and evolving dark energy scenarios.  Future CMB  in the microwave to far-infrared bands in the polarization and the amplitude, as the Polarized Radiation and Imaging Spectroscopy (PRISM) \cite{PRISM} and the very high precision measurements of the polarization of the microwave sky by the Cosmic Origins Explorer (CoRE) satellite \cite{CoRE} will improve the constraints of the dark sector \cite{SF-C-35}.

 The focus of this paper is on parametric reconstruction of some equation of states for DDE. The models chosen in this work have equation of states which behave quite diversity. The Chavelier-Polarski-Linder (CPL) and Feng-Shen-Li (FSL) models remain always below and above $ w=-1 $ respectively. On the other hand Power Law (PL) model has a crossing $ w=-1 $ behavior. The  CPL model  despite being widely used for long time, suffer from divergence as $z$ approach to -1. The alternative divergence-less model is FSL as $z$ approach to -1 instead of CPL proposed. The PL, on the other hand was proposed to resolve fine-tuning problem.  Thanks to its crossing behavior, the model  has a positive equation of state at early time. Leading it to naturally behave similar to  coupled dark energy-Dark matter models without introducing any coupling parameter.

 In this work,  from observational points of view, we will rely on the state-of-the-art observational data sets such as supernova type Ia (SNIa), baryonic acoustic oscillation (BAO), Hubble parameter, full sky CMB and also take into account the perturbations to examine the consistency of our models with Large Scale Structure (LSS) observables. We introduce a new geometrical measure, namely  number density of local maxima as discriminator between different DDE models. We also study perturbation of dark energy and its clustering again a potentially powerful tool to distinguish between various DDE models. Apart from the background field that drives the accelerating expansion of the universe, dark energy perturbations exist intrinsically. The dark energy clustering is relevant not only for structure formation at low redshifts $z\leq 1  $, where the energy density  dominates of the cosmic expansion, but also for early  structure formation when early dark energy exists. A reasonable DDE model with its  perturbation predicts that the dark energy can either cluster or produce voids. However, dark energy is smooth on small scales leading small scales structure such as galaxy formation and solar-system dynamics unaffected \cite{DEC-4}. We also investigate the implication of these DDE models for structure formation $ f\sigma_8 $ measurement, matter power spectrum and ISW effect and how they improve the Hubble tension.

The paper is organized as follows.  In Sec. 2 we describe  our phenomenological uncoupled DDE models . In Sec. 3  the theoretical framework of structure formation in present of DDE models will be discussed. These including evolution of scalar perturbation, Power spectrum of matter density, $\sigma_8$ and bias free parameter namely $f\sigma_8$ as a function of redshift will be included in this section. ISW are also given in Sec. 3.  Sec. 4 is devoted to the clustering of DDE models. Sec. 5 introduce  a new geometrical measure of the  number density of local maxima at the CMB maps in the presence of DDE models which may alter the CMB fluctuations even in very weak situation. Observational constraints based on JLA data sets, BAO, {\it Planck} TT, LSS will be given in Sec. 6. Results and discussion about our finding will be given in Sec. 7. We will give concluding remarks in Sec. 8.

\section{Dynamical Dark Energy  Models}\label{DDEmodel}
The nature of agent acceleration is still pending. To this end, various dark energy models as alternative models for cosmological constant have been introduced. A well-known category including a modification on constituent of universe is so-called  phenomenological model. This class is devoted to models with rolling fields \cite{MB-5}, modified kinetic terms \cite{47} and higher spin fields\cite{48}.  In phenomenological approaches, dark energy is mainly characterized by its equation of state  $ w=p / \rho $, speed of sound $ c_{s} ^{2}=\delta p / \delta\rho$ and anisotropy stress $\sigma$ \cite{Sf-4}. It turns out that DDE models can be distinguished from cosmological constant by time dependence equation of state and non-zero sound speed .
The number of free parameters of DDE models depend on theoretical and phenomenological approaches  are utilized for construction. However, it was revealed that  to achieve decorrelate compression and to avoid crippling limitation, we need at least  three free parameters \cite{MB-3}.

In this section, we will  briefly explain three dark energy models
considered for examining clustering of dark energy in the context of
perturbation theory.  Such dark energy models can be modeled by a
barotropic perfect fluid, namely, its pressure just can be
considered as a function of density with a proper equation of state
depending on redshift. As following, the DDE models have various
behavior versus redshift in comparison with $\Lambda$CDM. PL has
Phantom-crossing manner which leads similar behavior to dark matter
at early time. The FSL model remains always higher than $-1$, while
CPL remains less than $-1$. Subsequently, one can asses such
classification in the clustering behavior (see Fig. \ref{fig:w(z)}).

\subsection{CPL model}
The first order covariant expansion for equation of state is $  w(z)= w_{0}+w_{1} z$ which is unstable at high redshift. Hence to resolve this problem, Chavelier-Polarski-Linder model (CPL) has been proposed equation of state in the form of \cite{Chevallier:2000qy,M-1}:
\begin{equation} \label{eq:wcpl}
w_{\rm CPL}(z)= w_{0}+w_{1}\dfrac{z}{1+z}
\end{equation}
where $w_0$ and $w_1$ are the parameters of equation of state at
present time and derivative of equation of state with respect to
scale factor, respectively. Several advantage of this model are
explored such as manageable two dimensional phase space, bounded
behavior at high redshift and reconstruction of many scalar field
equation of states with high accuracy. In addition, it  possesses
fine physical interpretation. Beside mentioned advantages, its
equation of state diverge at $z=-1$ \cite{MB-2}. Recent
observational consistency test based on supernova revealed that
there is no significant evidence of any deviation from linearity
from with respect to scale factor  \cite{Salzano:2012zp}.

\subsection{FSL model}
As mentioned before, CPL model suffers divergence problem when redshift approaches to $z\to-1$ leading to non-physical future. Hence  Feng-Shen-Li (FSL) proposed a model to eliminate this problem with non-infinite limit as \citep{M-5}
\begin{equation}\label{eq:wfsl}
w_{\rm FSL}(z)=w_{0}+w_{1}\frac{z}{1+z^{2}}
\end{equation}
 The free parameters are similar to CPL model.

\subsection{PL model}
 Another model utilized for evaluation of dark energy is power-law (PL) parameterized model proposed in order to solve fine tuning problem as \cite{M-3}:
\begin{equation}\label{eq:wpl}
%w_{\rm PL}(a) = w_{0}a^{\alpha}[1 + \ln\left(a^{\alpha} \right)]
w_{\rm PL}(z) = \frac{w_{0}}{(1+z)^{\alpha}}[1 -\alpha \ln\left(1+z \right)]
\end{equation}
in which $w_{0} $ and $\alpha $ are model's free parameters. The ratio of the dark energy density to the matter density is not sensitive to value of model's free parameters and asymptotically goes to zero at early epoch.
Hence dark energy does not need to be fine tuned at $ t \rightarrow 0 $. PL model also solved age of old stars problem which is called cosmic age crisis \cite{M-3} (for full review of the cosmic age see \cite{julian67}). Another advantage of the model is that for scale factor in range of $ a \propto e^{\frac{1}{\alpha}} $ the sign of equation of state changes to positive value, so it can be a candidate for unified dark energy-dark matter models scenario.

\begin{figure}[t]
\centering
        \includegraphics[width=0.5\textwidth]{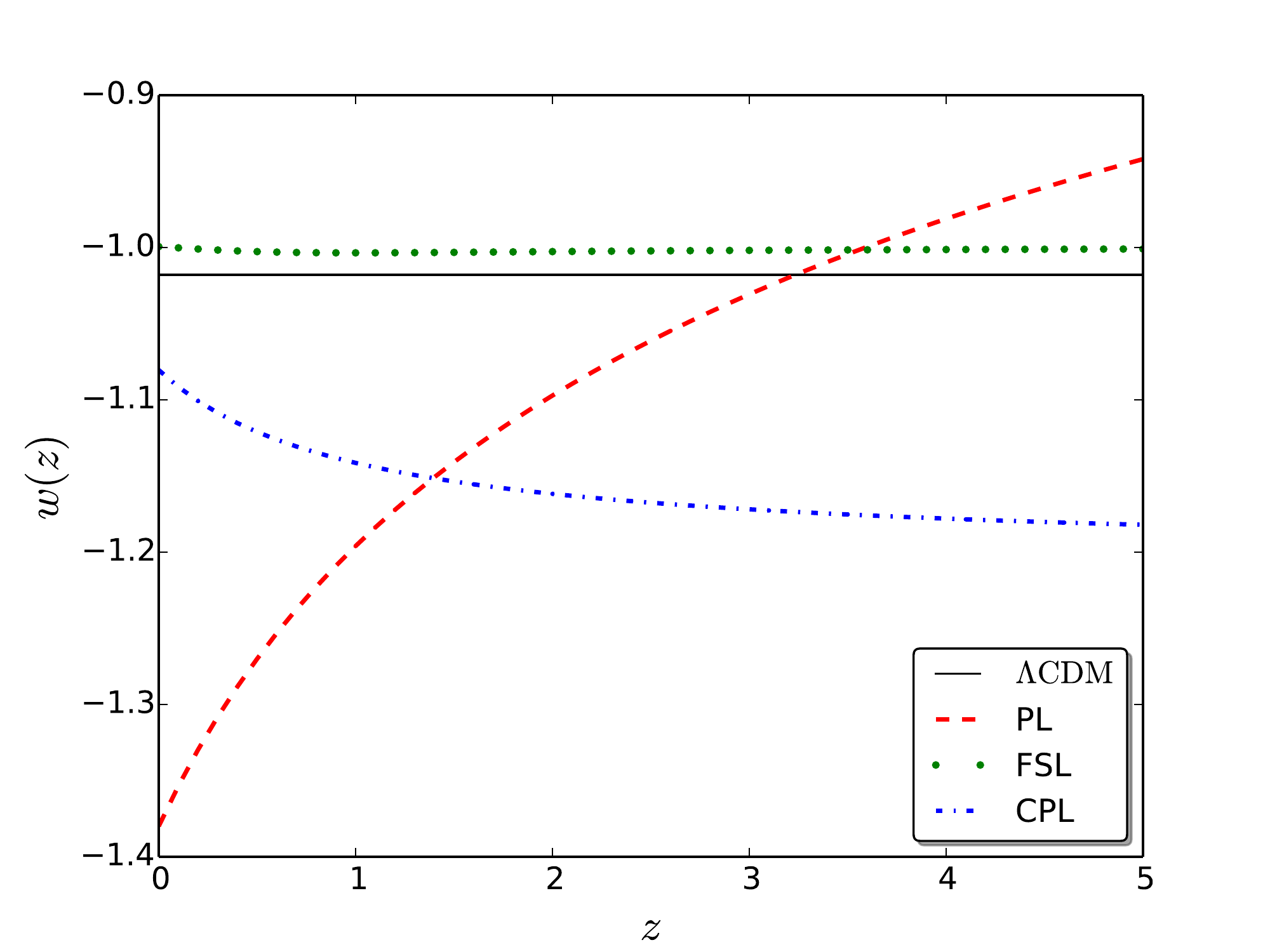}
     \caption{The equation of states of the DDE models as function of redshift.  The value of free parameters set to their best fit values determined by observational constraints using by SNIa+BAO+HST+$Planck$ TT+LSS (see section \ref{s:constrains}). }\label{fig:w(z)}
\end{figure}

In Fig. \ref{fig:w(z)}, we plot the equation of states associated
with  the mentioned models. We used best fit values for
corresponding free parameters based on different observational data
sets to illustrate the behavior of  equation of states (see section
\ref{s:constrains}). The PL equation of state  crosses  the
cosmological constant at late time and it almost behaves as matter
component at early epoch. In the next section we will rely on the
hydrodynamical linear perturbation theory to assess inhomogeneities
associated with energy constituents.

\section{Structure Formation in DDE I: Standard approach}
In this section, we will explain useful theoretical framework to elaborate perturbative dynamics for components universe. First order perturbation on various components and associated velocity dispersion will be discussed in this section. We ignore any interaction between dark sectors of the universe.

\subsection{Evolution of Scalar Perturbation}
In order to realize the evolution of perturbation in cold dark matter (CDM) and in a typical  DDE component in the framework of perturbation approach, one can decompose the Einstein tensor, $G_{\mu\nu}$, and the energy-momentum tensor, $T_{\mu\nu}$, into background and perturbed parts. A general functional form for $T_{\mu\nu}$ considered as:
\begin{eqnarray}
T_{\mu \nu}=T_{\mu \nu}^{\rm{m}}+T_{\mu \nu}^{\rm{rad}}+ T^{\rm {DE}}_{\mu \nu},
\end{eqnarray}

here $T^{\rm{m}}_{\mu \nu}$, $T^{\rm{rad}}_{\mu \nu}$ and $T^{\rm {DE}}_{\mu \nu}$ are the energy-momentum tensors of the matter, radiation and dark energy, respectively. Tensor $T^{\rm {DE}}_{\mu\nu}$ includes other sources of gravity such as scalar fields \cite{Caldwell:1999ew}. In this paper for the dark energy part, we assume  that $w_{\rm DE}$ takes one of DDE model's equation of state  which introduced in previous section.
The generalized equation of state reads as:
\begin{equation}
\bar{w}_{i}(a)\equiv  \frac{\int_1^a w_{i}(a') d\ln a'}{\int_1^a d\ln a'},
\end{equation}
 here $w_i$ is equation of state of $i$th component. As an illustration for a given DDE models,  $w_{\rm DE}$ is represented by Eqs. (\ref{eq:wcpl}), (\ref{eq:wpl}) and (\ref{eq:wfsl}). The functional form of generalized equation of state for CPL, FSL and PL are:
 \begin{eqnarray}
\bar{w}_{\rm CPL}(a)&=&\omega_{1}+(\omega_{0}-\omega_{1})\frac{a-1}{\ln(a)},\\
\bar{w}_{\rm FSL}(a)&=&\omega_0-\omega_1 \frac{\arctan\left(\frac{1-a}{a}\right)}{\ln(a)},\\
\bar{w}_{\rm PL}(a)&=& \omega_{0} a^{\alpha}.
\end{eqnarray}
Considering convention by line element in so-called Newtonian or longitudinal gauge, we get:
\begin{equation}
ds^{2}=a^{2}(\eta)\left[ -(1+2\psi)d\eta^{2}+(1-2\phi)dx^{i}dx^{j}\right],
\end{equation}
 where $\eta$ is conformal time, $d\eta=dt/a$. Also $ \phi $ and $ \psi $ are space-like and time-like variables, respectively describing scalar metric perturbations and  known as gauge invariant Bardeen potentials \cite{PER-1}. Gauge fixing influences the perturbations especially on scales larger than Hubble horizon $ k\leq aH $, while on much smaller scales the choice of gauge is less important and observables are independent of gauge choice \cite{SF-9}.

 Hence  Einstein perturbed equations and perturbed continuity fluid equation read as:
\begin{eqnarray}\label{eq:1}
 k^{2} \phi+ 3 \mathcal{H}(\phi^{\prime \prime} - \mathcal{H}\psi) &=&4\pi G a^{2} \rho \delta , \\
\label{eq:2}
k^{2} \phi^\prime - \mathcal{H} \psi &=& 4\pi G a^{2} (1+\omega)\phi \theta , \\
\phi^{\prime \prime} + 2\mathcal{H}\phi^\prime +\mathcal{H}\psi^\prime -(\mathcal{H}^{2}+ 2\mathcal{H}^\prime)\psi &=& 4\pi G a^{2} c_{s}^{2}\rho \delta  ,\\
\delta^\prime + 3 \mathcal{H} (c_{s}^{2}-w)\delta &=& -(1+ w)(\theta + 3 \phi^\prime) , \\
\theta^\prime +\left[ \mathcal{H}(1- 3 w)+\dfrac{w^\prime}{1+ w }\right] \theta &=& k^{2}(\dfrac{c_{s}^{2}}{1+ w} \delta + \psi ) .\label{eq:5}
\end{eqnarray}
where the prime denotes derivative with  respect to conformal time, $\delta \equiv\delta \rho/\rho$ is so-called density contrast, $ w $ is equation of state of fluid, $ c_{s} ^{2} $ is sound speed  and $ \theta =i \textbf{k}. \textbf{v} $ is velocity dispersion. We also ignore non-zero anisotropic stress leading to anisotropy, consequently $\psi=\phi$. If we consider Eqs \ref{eq:1} and \ref{eq:2} we can obtain:
\begin{equation}
k^{2}\phi =-4 \pi G a^{2} \rho \Delta ,
\end{equation}
here
\begin{equation}\label{eq:Deltaobs}
\Delta\equiv\frac{3(1+\omega)}{k^{2}} \left(\frac{\dot{a}}{a}\right) \theta +\delta ,
\end{equation}
which $ \Delta $ is a physical observable. When we are interested in evolution of small scale (inside the Hubble radius), the Newtonian potential $ \psi $ and $ \phi $ do not  vary in time along the matter era, therefore, the quasi static approximation is conceivable and  those terms including $k^{2}/a^{2} $ become dominant. Apply the quasi static approximation for  \ref{eq:1} leads to:
\begin{equation}
k^{2}\phi =-4 \pi G a^{2} \rho \delta .
\end{equation}

All of the coupled equations mentioned before should be solved with proper initial conditions in order to achieve evolution of various types of perturbations in an expanding universe.

\subsection{Power Spectrum and $ \sigma_{8} $}
Power spectrum describes density contrast in the universe as a function of scale in the Fourier space. It is a relevant observable quantity corresponding to Fourier transform of two-point correlation function of underlying fluctuations. This quantity describes scale dependency of fluctuations. The most popular assumption is that the primordial fluctuations has been distributed according to homogeneous Gaussian random fields which comes from simple model of inflation \cite{POS-M-2}. According to mentioned assumption, all statistical information is encoded in two-point correlation function or equivalently in the power spectrum.
According to linear Perturbation theory, Structures grow on large scale if local gravity wins the competition against the cosmic expansion. On the other hand, on the small scales, gravitational collapse is non-linear process and can only  be fully addressed with N-body simulation.

The cold dark matter power spectrum is defined as $P(k,a)\equiv k^{n_s} T^2(k) D^{2}(a)$. In which, $D(a)$ is the linear growth factor and it is independent of scale.  Also $T(k)$ is the cold dark matter transfer function \cite{ma99,dodelson,bardeen86,Pouri:2014nta}. Recent analysis based on $Planck$ data demonstrated that $n_s \simeq 0.9655\pm0.0062$ \cite{Ade:2015xua}. The root-mean-square of fluctuations of the linear density field on mass scale $M$ is:
\begin{equation}\label{eq:sigma-rms}
\sigma(M, z) = \left[\frac{1}{2 \pi^2} \int_0^\infty k^2 P(k,z) W^2(kR) dk \right]^{1/2} ,
\end{equation}
where $W(kR) = \frac{3(\sin kR - kR \cos kR)}{(kR)^3} $ and $R=(3M/4 \pi \rho_{\rm m})^{1/3}$. The root-mean-square mass fluctuation field on $R_8 = 8 h^{-1}$ Mpc is called $\sigma_8(a)\equiv \sigma (R_8,a)$. It is worth noting that model independency is an important property associated with a typical observable quantity which causes to infer reliable results.The mentioned quantity is thought as almost model-dependent and particularly it depends on galaxy density bias \cite{song09}. To get rid of this discrepancy, we turn to define another alternative composition of observable quantities. To this end, according to linear growth factor, $D(a)$,  the growth rate is defined by:
\begin{equation}
f(a) \equiv \frac{d \ln D(a)}{d \ln a} .
\end{equation}
Most growth rate measurements are based on peculiar velocities obtained from Redshift Space Distortion (RSD) measurements \cite{Kaiser:1987qv}. In principle, comparison between transverse against line of sight anisotropies influenced by peculiar motion in the redshift space clustering of galaxies yields observational constraints on proper quantities coming from linear perturbation theory. Galaxy redshift surveys provide measurements of perturbations in terms of galaxy density $\delta_g$, which are related to matter perturbations through the bias parameter $b$ as $\delta_g=b \delta_m$. There is a robust combination, namely  $f \sigma_8(z)\equiv f(z) \sigma_8(z)$ which is independent of the bias factor, and could be achieved utilizing  weak lensing and RSD \cite{song09,Nesseris:2017vor}. However this parameter has a degeneracy with Alcock-Paczynski (AP) effect. In this paper we use the  observable values for $f\sigma_8$ in order to put constraints on free parameters of DDE models.

\subsection{Integrated Sachs-Wolfe Effect}
In this section we study the systematic behavior of temperature
anisotropies power spectrum in the presence of dynamical dark energy
models introduced in section \ref{DDEmodel}. However the main
contribution of dark energy is for late time but for DDE models, it
is not trivial to ignore corresponding effects on temperature
anisotropy power spectrum due to late and early Integrated
Sachs-Wolfe $(ISW)$ effects. Temperature anisotropy power spectrum
is represented by temperature fluctuations correlation function
expanded in spherical harmonics \cite{ISW-3}:
\begin{equation}
C_{\ell}=4\pi \int \dfrac{dk}{k} P(k)|\mathcal{D}_{\ell}(k,\eta_{0})\vert^{2} .
\end{equation}
$P(k)$ is primordial power spectrum and $ \mathcal{D}_{\ell}(k,\eta_{0}) $ gives transfer function for each $\ell$:
\begin{equation}
\mathcal{D}_{\ell}(k,\eta_{0})= \mathcal{D}_{\ell}^{LSS}(k)+\mathcal{D}_{\ell}^{ISW}(k),
\end{equation}
where $ \mathcal{D}_{\ell}^{LSS}(k) $ represents ordinary Sachs-Wolfe effect and $  \mathcal{D}_{\ell}^{ISW}(k) $ includes the contribution of variation of potential along of line of sight (Integrated Sachs-Wolfe effect) \cite{sw67,hu95}:
\begin{equation}
\mathcal{D}_{\ell}^{ISW}(k)=2 \int d\eta  e^{-\tau}  \Phi^\prime j_{\ell}(k(\eta - \eta _0)) .
\end{equation}
Hence, $ \tau $ is optical depth due to photon scattering along the line of sight and $ j_{\ell} $ is spherical Bessel function.
From physical point of view, when CMB photons traveling from last scattering surface (LSS) to observer are entering (leaving) high dense regions and for low dense, associated regions receives blue shifted (red shifted).
 If gravitational potential $ \phi $ evolves during a photon crossing the different regions when dark energy  exists, the both effect will not cancel out each other and final energy of photon varies \cite{ISW-2}. Accordingly, in matter dominated era, $ \phi $ remains constant and we have no $ISW$ effect. However, when dark energy dominates, $ \phi $ is no longer constant and $ISW$ effect generates secondary anisotropy on CMB map. For this reason, $ISW$ effect is a robust method for distinguishing and comparing the various variable dark energy models. Solving the coupled perturbed Einstein equations (Eqs. (\ref{eq:1})-(\ref{eq:5})), one can determine evolution of gravitational potential $ \phi $ in linear regime for various DDE models hence the contribution on temperature power spectrum beyond zero-order will be achieved.
 The ISW effect is not only sourced by local structure but also secondary anisotropy could be produced by non-linear evolution of gravitational collapse of small scale in cluster and super cluster but pertinent scale is correspond to angular 5-10 arcmin, much smaller than those associated by ISW effect\cite{ISW-C-3}.

\begin{figure}[tbp]
\centering
\includegraphics[width=.45\textwidth,origin=c,angle=0]{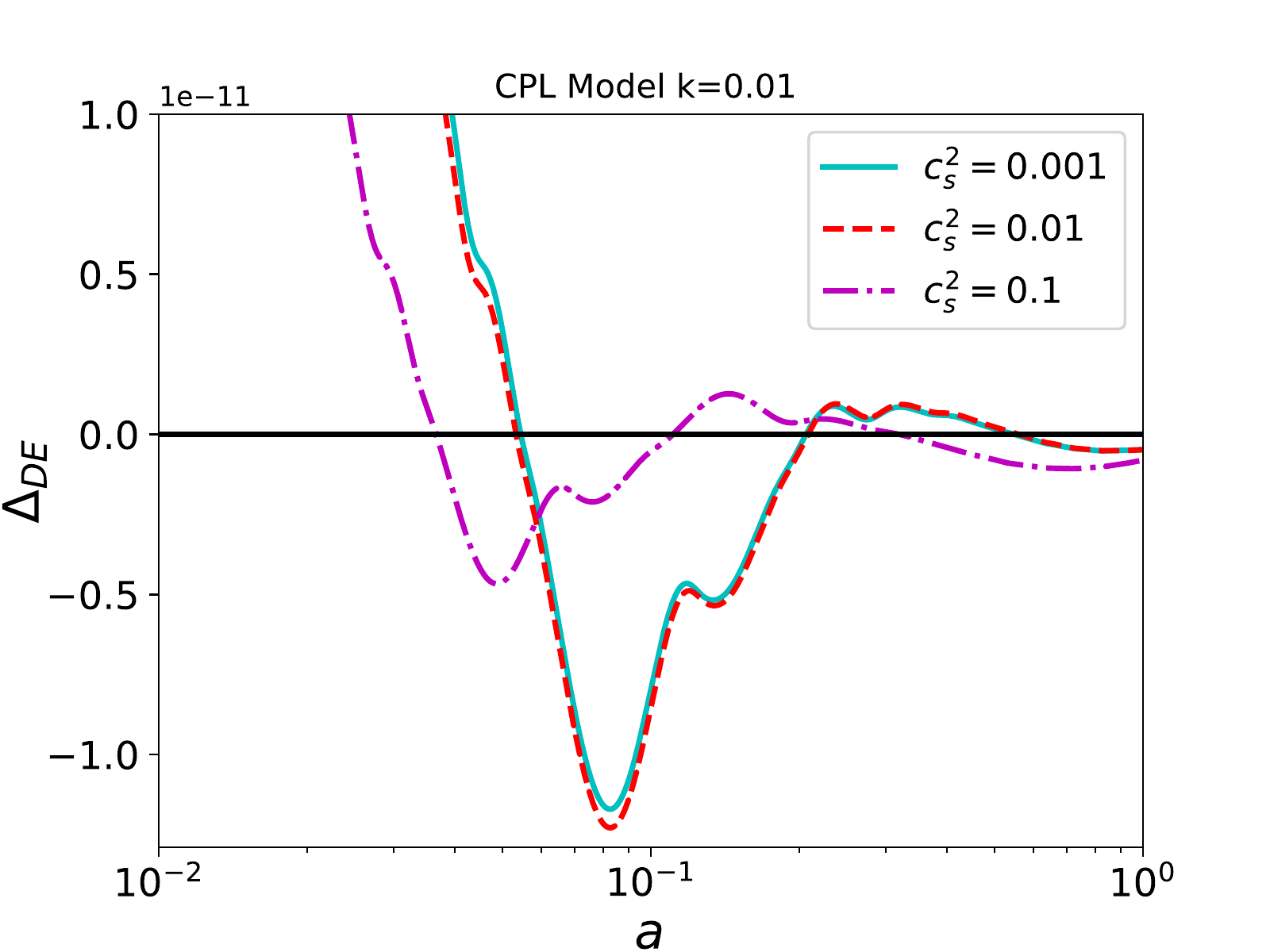}
\includegraphics[width=.45\textwidth,origin=c,angle=0]{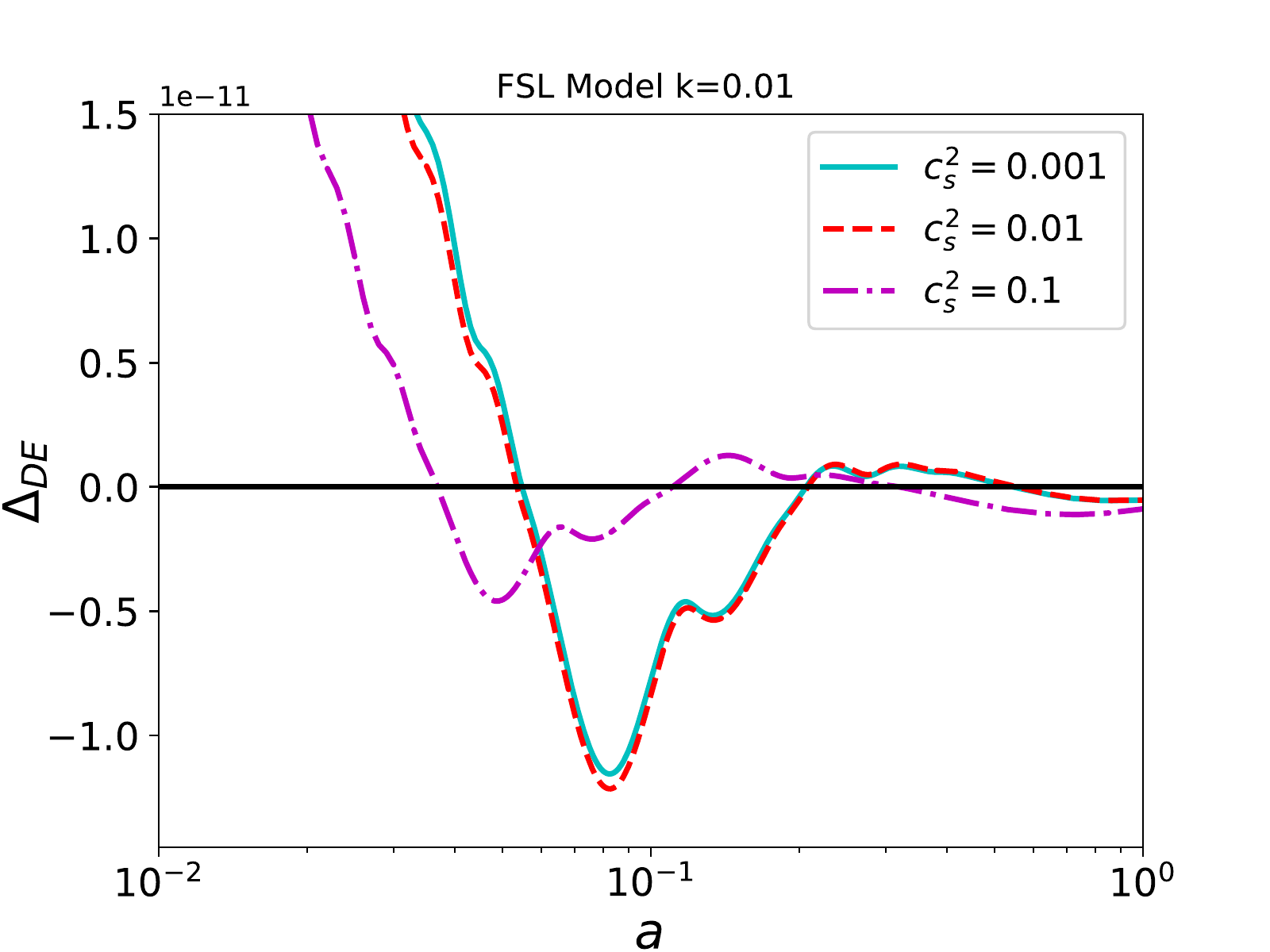}
\includegraphics[width=0.5\textwidth]{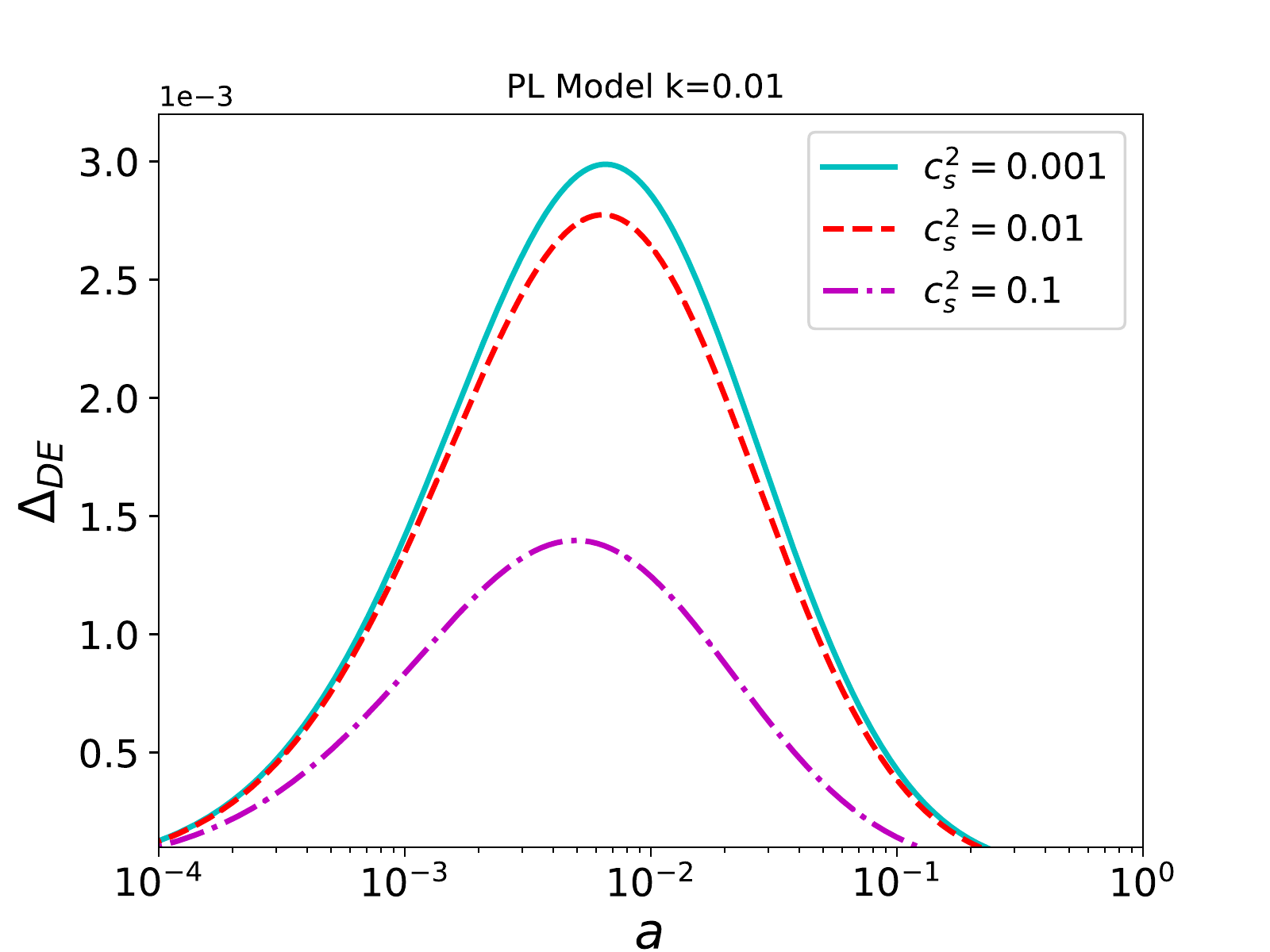}

\caption{\label{fig:DEC} Clustering of dark energy at
$k=0.01Mpc^{-1}$ affected by corresponding sound speed. Upper left
panel indicates $\Delta_{\rm DE}$ for the CPL, upper right panel
corresponds to $\Delta_{\rm DE}$ for the  FSL, while lower panel is
for $\Delta_{\rm DE}$ of the PL dark energy model. For PL model,
clustering adopts only positive value associating with over-density
while for CPL and FSL, we find negative and positive value of
generalized density contrast. At the early epoch the value of
clustering goes to its initial conditions. We take best fit values
for free parameters determined by combining SNIa+ BAO+ HST+$Planck$
TT+LSS data sets. }
\end{figure}

\section{Structure Formation in DDE II: Clustering of Cold Dynamical Dark Energy} \label{DDE clustering}
An important feature of dynamical dark energy is spatial perturbation, providing an independent key to investigate the nature of dark energy from equation of state and sound speed. The sound speed carries information about internal degree of freedom and low sound speed enhances the spatial variation of dark energy, giving rise to inhomogeneities or clusters. Clustering is a benchmark of dark energy perturbation which can affect on matter power spectrum and large scale clustering. It is investigated that  a void on matter was seen to produce by local DDE overdensity and suggested DDE clustering might be relevant strong when matter perturbation goes non-linear \cite{DEC-2}. In order to have considerable effect on matter power spectrum due to clustering of a typical dark energy, two necessity conditions should be satisfied \cite{DEC-1}:  dark energy equation of state must deviate from $-1$. All perturbations vanish  as $ 1+w\rightarrow 0$, regardless of  $ c_{a}^{2} $.
Secondly, the sound horizon $ c_{s} H^{-1}$ of dark energy perturbation must be well within the Hubble scale that required $ c_{s}^{2} \ll 1$ which considered as Cold Dark Energy (CDE).

 Exact value of $ c_{a}^{2} $ is hard to determine and degeneracy with other parameters come into play.
In general case, sound speed is defined by:
\begin{equation}\label{eq:sound1}
c_{s}^{2} \equiv\frac{\delta P}{\delta \rho}=c_{s}^{2(A)}+c_{s}^{2(NA)} .
\end{equation}
where adiabatic sound speed, $ c_{s}^{2(A)}$, is purely determine by equation of state:
\begin{equation}\label{eq:sound}
c_{s}^{2(A)} \equiv\dfrac{ p\prime}{\rho \prime}= w_{\rm DE}- \dfrac{w \prime_{\rm DE}}{3{\mathcal{H}}(1+w_{\rm DE})}
\end{equation}
where prime denote derivatives with respect to
conformal time and $\mathcal{H}$ is the Hubble constant
with respect to conformal time. Also $c_{s}^{2(NA)}$ represents non-adiabatic term.  In general case, the pressure may depend on internal degree of freedom of the underlying fluid.

Supposing the dark energy behaves strictly as adiabatic fluid, consequently, sound speed is mainly affected by adiabatic term. For $c_s^2<0$, instability appears and to resolve this discrepancy, one can consider some classes of dynamical dark energy including other components which effectively causes to have $w_{eff}>0$ \cite{DEC-6}.  Another approach is supposing the non-adiabatic term in Eq. (\ref{eq:sound1}) to be survived.

 In order to have stable clustering, the sound speed should adopt in the range of very small value to unity. Here in this paper we take $ c_{s}^{2}=0.1$, $ c_{s}^{2}=0.01$ and $ c_{s}^{2}=0.001$ for clustering of dark energy at large scale $k=0.01 h/Mpc $. In Fig. \ref{fig:DEC}, we illustrate  clustering of dark energy models as a function of scale factor for $k=0.01 h/Mpc$ for various values of sound speed.

\section{Data Analysis I: Number of local maxima peaks }
\begin{figure}[t]
\centering
        \includegraphics[width=0.5\textwidth]{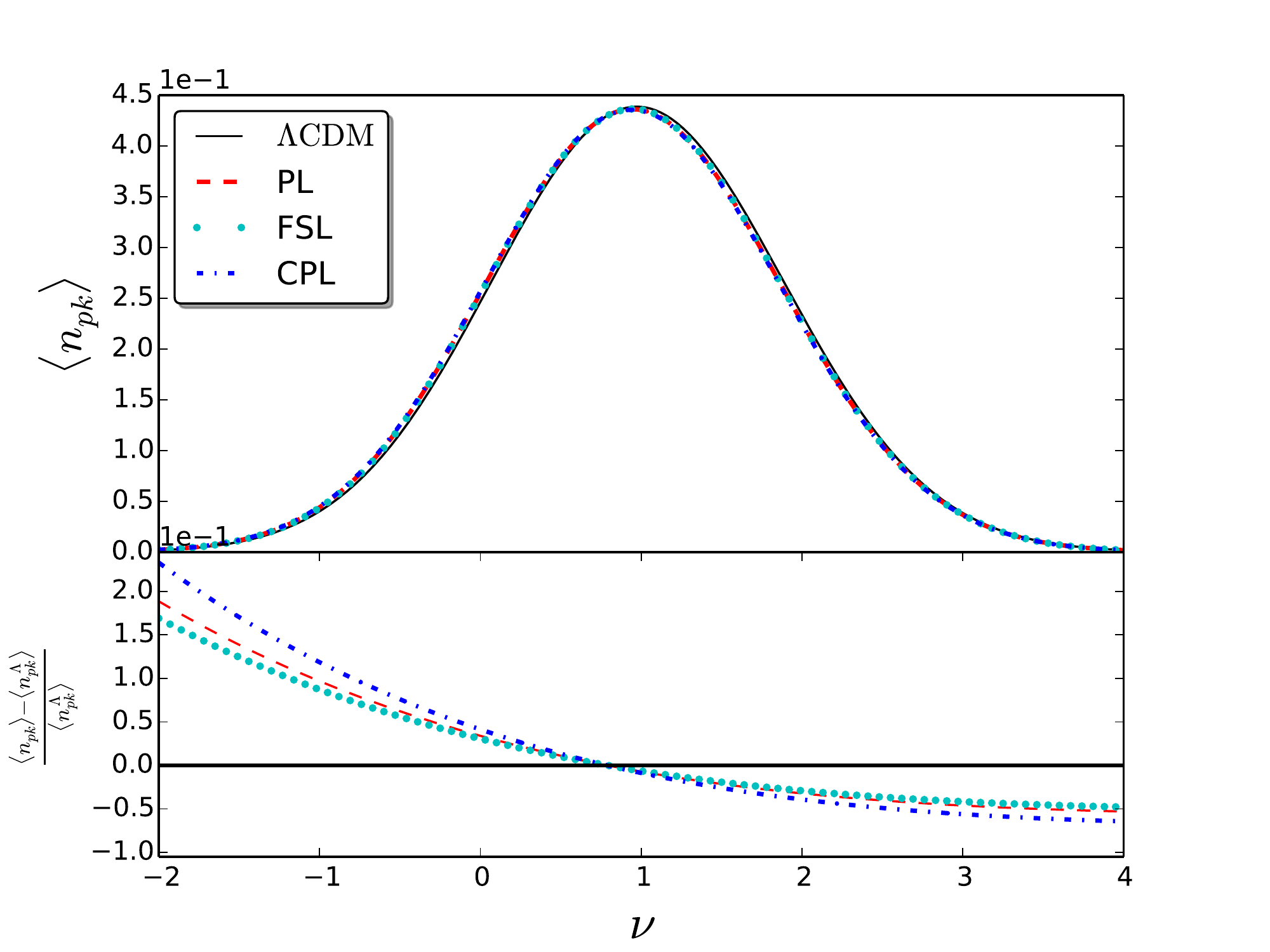}
     \caption{ The number density of peaks for full sky CMB map in the presence of dynamical dark energy models. We consider the best fit values for free parameters of models from joint analysis SNIa+BAO+HST+$Planck$ TT+LSS. }\label{fig:ND}
\end{figure}

One of our aims is finding a proper way to distinguish between different DDE models. Many measures have been proposed ranging from  geometrical and topological  approaches to classical aspects \cite{jain03,taylor07,Kitching08,Chenxiaoji:2014mxa,Shirasaki:2016twn,Fang:2017daj}. For example  four minkowski functionals of morphological properties used of large scale structure were used to distinguish modified gravity models from general relativity \cite{Fang:2017daj}. Similarly, one can apply method base on cluster number counts method and number density of peaks trough to discriminate between models of DDE.  To this end, we need to determine various order of for 2 dimensional field, given by:
\begin{equation}
\sigma^{2}_{m} \equiv  \langle \triangledown^{m}\delta_{T} \triangledown^{m}\delta_{T} \rangle = \frac{1}{2\pi}\int dk k^{2m} P_{TT}(\vert k\vert) W^{2}(kR).
\end{equation}
Hence $P_{TT}(k)$ is the power spectrum and $W$ stands for any smoothing function and $R$ is the smoothing scale. These parameter for a full sky CMB map are:
\begin{equation}
\sigma^{2}_{m}\equiv \sum_{\ell} \frac{(2 \ell +1)}{4\pi}[\ell (\ell+1)]^{m} C_{\ell}^{TT} W_{\ell}^{2},
\end{equation}
$C_{\ell}^{TT}$ represent the  power spectrum for the full sky.  $W_{\ell} $ is smoothing kernel associated with beam transfer function which is written by: $W_{\ell} = \exp(-\theta_{\rm beam}\ell(\ell+1)/2)$ and $\theta_{\rm beam}=\theta_{\rm FWHM}/\sqrt{8\ln(2)}$ \cite{Bond and efstathiou 1987,ravi99,Hikage:2006fe}. In the flat sky approximation one can write:
\begin{equation}
\frac{\ell (\ell+1)C_{\ell}^{TT}}{2\pi} \backsim \frac{(360)^{2}}{2 \pi} \vert k\vert^{2} P_{TT}(k) .
\end{equation}
By imposing the condition to have peaks above a given threshold, we find that:
\begin{eqnarray}
\langle n_{p}(\vartheta;r)\rangle&=& \langle \Theta(\delta_{T}-\vartheta \sigma_{0})\delta_{D}(\eta_{\phi}) \delta_{D}(\eta_{\phi}) |{\rm Det}\xi| \rangle  \nonumber\\
&=& \int_{\rm conditions} \textit{P} ( A_{\mu}| \delta_{T} \geqslant \vartheta \sigma_{0}; \eta _{\phi} =\eta_{\phi} =0) |{\rm Det}\xi| d A_{\mu} ,
\end{eqnarray}
where $ \Theta$ is the unit step function. The subscript
"conditions" represents additional considerations for having local
maxima.  The number density of peaks for a purely homogeneous
Gaussian CMB map in the rang of $ [ \vartheta , \vartheta
+d\vartheta] $ becomes\cite{Bond and efstathiou 1987}:
\begin{equation}
n_{p}(\vartheta)=\frac{N_{\rm pix,tot}}{(2\pi)^{3/2} \gamma ^{2}} e^{\frac{-\vartheta^{2}}{2}} G(\Gamma, \Gamma\vartheta) ,
\end{equation}
where
\begin{eqnarray}
G(\Gamma, \Gamma\vartheta)&=&(\Gamma^{2}\vartheta^{2}-\Gamma^{2})\left\{1-\frac{1}{2}{\rm erfc}\left[\frac{\Gamma\vartheta}{\sqrt{2(1-\Gamma^2)}}\right]\right\}+\Gamma\vartheta(1-\Gamma^{2})\frac{\exp \left({-\frac{\Gamma^{2}\vartheta^{2}}{2(1-\Gamma^{2})}}\right)}{\sqrt{2\pi(1-\Gamma^2)}}\nonumber\\
&&+\frac{\exp\left({-\frac{\Gamma^{2}\vartheta^{2}}{3-2\Gamma^{2}}}\right)}{ \sqrt{3-2\Gamma^{2}}}\left\{1-\frac{1}{2} {\rm erfc} \left[\frac{\Gamma\vartheta}{\sqrt{2(1-\Gamma^{2})(3-2\Gamma^{2})}}\right]\right\} .
\end{eqnarray}
Hence erfc(:) is the standard complementary  error function, and the parameters  $ \Gamma $ and $ \gamma $ are defined as $ \Gamma\equiv\frac{\sigma_{1}^{2}}{\sigma_{0}\sigma_{1}} $ and $ \gamma\equiv \sqrt{2}\frac{\sigma_{1}}{\sigma_{2}} $. Also $N_{\rm pix,tot}$ is the total number of pixel in a given map. Fig. \ref{fig:ND}  depicts the theoretical prediction of number density of peaks at a given threshold for three DDE models and cosmological constant.  One can use local extrema for high negative or for high positive thresholds where the sensitivity of measure with respect to various models is reasonable.

\section{Data Analysis II: Observational constraints} \label{s:constrains}
In the current section, we will rely on the most recent observational data sets to revisit the observational consistency of underlying dynamical dark energy models explained in section \ref{DDEmodel}. Accordingly, we will find the best fit values for associated free parameters in order to assess not only the impact of model at the background level but also to determine the clustering contribution of dark energy.  Generally, following observables are utilized  for further evaluation:  \\
Geometric methods including various cosmological distance measures and dynamical methods which is mainly determined by Hubble parameter. Mentioned probes are mostly known as background contributions and to find more accurate identification, we can go beyond zero-order approximation and relying on higher order represented by evolution of perturbations. To this end, linear growth of density perturbations will be exploited to constrain  free parameters of the models.  We also examine evolution of gravitational potential, causing to Integrated Sachs-Wolfe (ISW) effect and also matter power spectrum and $ f\sigma_8 $ of models obtained. We will compute CMB power spectrum in the presence of DDE model neglecting the contribution of DDE clustering.  Additional probes such as cross-correlation of CMB and large scale structures and Weak lensing will be left for future researches.  In this paper we take into account the background expansion indicators such as distance modulus of Supernovae Type Ia and the Hubble parameter, Baryon acoustic oscillations (BAO), full sky CMB and $ f\sigma_8 $ data as flowing, our parameter space is:
\begin{equation} \label{eq:parameter-space}
\{\mathcal{P}\} \equiv \left\lbrace \Omega_{\rm b} h^2 , \Omega_{\rm c} h^2 , \Theta, \tau , w_0 , w_1 , \alpha, n_s , \ln \left[ 10^{10} \mathcal{A}_s \right] \right\rbrace .
\end{equation}
Hence, $\Omega_{\rm b} h^2$ and $\Omega_{\rm c} h^2$ are the physical baryon and cold dark matter densities, $\Theta$ is the ratio (multiplied by 100) of the sound horizon to angular diameter distance at decoupling, $\tau$ is the optical depth to re-ionization, $w_0$ and $w_1$ and $\alpha$ are  free parameters of dark energy equation of states, $n_s$ is the scalar spectral index and $\mathcal{A}_s$ is defined as the amplitude of the initial power spectrum. The pivot scale of the initial scalar power spectrum we used here is $k_p=0.05 Mpc^{-1}$ and we have assumed purely adiabatic initial condition. We have also supposed our Universe is flat, so $\Omega_{tot} =
\Omega_{\rm c}+\Omega_{\rm b}+\Omega_{\rm rad}+\Omega_{\rm DE}=1$. The priors for free parameters of  models have been mentioned in Table \ref{prior}.

\begin{table}[]
\begin{center}
\centering
\begin{tabular}{llccc}
\hline
\hline Parameter & Prior & Shape of PDF \\
\hline $\Omega_{tot}$ &1.000& Fixed&\\
 $\Omega_bh^2$ &$[0.005-0.100]$& Top-Hat&\\
 $\Omega_c h^2$ &$[0.001-0.990]$& Top-Hat&\\
 $\alpha$ &$[0.00-0.20]$& Top-Hat&\\
 $w_0$ &$[-1.2- -0.9]$& Top-Hat&\\
 $w_1$ &$[-0.2- 0.2]$& Top-Hat&\\
 $H_0$ &$[40.0-100.0]$& Top-Hat&\\
 $\tau$ &$[0.01-0.80]$& Top-Hat&\\
 $n_s$ &$[0.800-1.200]$& Top-Hat&\\
 $\ln(10^{10} \mathcal{A}_s)$ &$[2.000-4.000]$& Top-Hat&\\
\hline
\hline
\end{tabular}
\caption{\label{prior} Priors on parameter space, used in the posterior analysis. }
\end{center}
\end{table}

\subsection{Luminosity distance }
Type-Ia supernovae (SNIa) are among the most important probes of the Universe expansion and  the main evidence for accelerating expansion epoch \cite{Riess:1998cb,Perlmutter:1998np}. Supernovae are considered as "standardizable candles" providing a measure for determining luminosity distance as a function of redshift.  SNIa is categorized in cataclysmic variable stars which are created  due to explosion of a white dwarf star and therefore its thermonuclear explosion is well known. Some of most important surveys of SNIa are: Higher-Z Team \cite{ries04,ries07}, the Supernova Legacy Survey (SNLS) \cite{ast06,bau08,reg09,guy10}, the ESSENCE \cite{mik07,wood07}, the Nearby Supernova Factory (NSF) \cite{cop06,scal09}, the Carnegie Supernova Project (CSP) \cite{fol10,fol101}, the Lick Observatory Supernova Search (LOSS) \cite{lea10,li10}, the Sloan Digital Sky Survey (SDSS) SN Survey \cite{holtz08,kess09}, Union2.1 SNIa dataset \cite{Suzuki12,Cao:2014jza}. Recently Joint Light-curve Analysis (JLA) catalogue was produced from SNLS and SDSS SNIa compilation \cite{Betoule14}.  In this paper we  consider SNIa observation data sets using JLA dataset including 740 SNIa in the redshift range of $z\in [0.01,1.30]$ \cite{Betoule14}.

However, the absolute luminosity of SNIa is considered uncertain and is marginalized out, which also removes any constraints on $H_0$. On the other hand, SNIa observations include low redshift data and effectively cover late-time expansion. So it could provide good constraints on models parameters.
In practice direct observation of SNIa is given by corresponding distance modulus:
\begin{eqnarray}
\mu (z;\{\mathcal{P}\}) \equiv m - M = 5\log \left( {\frac{{{d_L}(z;\{\mathcal{P}\})}}{{\rm{Mpc}}}} \right) + 25 ,
\end{eqnarray}
where $m$ and $M$ are apparent and absolute magnitude, respectively. For the spatially flat universe the luminosity distance defined in above equation reads as:
\begin{eqnarray}
{d_L}(z;\{\mathcal{P}\}) = \frac{c}{{{H_0}}}(1 + z)\mathop \int_0^z \frac{dz'}{\mathcal{H}(z';\{\mathcal{P}\})} ,
\end{eqnarray}
here $\mathcal{H}\equiv H/H_0$ and $H$ is Hubble parameter.  In order to compare observational data set which predicted by model we utilize likelihood function with following $\chi^2$ form:
\begin{eqnarray}
\chi^2_{SNIa}\equiv \Delta\mu^{\dag}\cdot\mathcal{C}^{-1}_{SNIa}\cdot\Delta\mu ,
\end{eqnarray}
where $ \Delta\mu\equiv \mu_{obs}(z)-\mu(z;\{\Theta\})$ and $\mathcal{C}_{SNIa}$ is covariance matrix of SNIa data sets. $\mu_{obs}(z)$ is observed distance modulus for a SNIa located at redshift $z$. Marginalizing over $H_0$ as a nuisance parameter yields \cite{Ade:2015rim,Li:2010da}
\begin{eqnarray}
\chi^2_{SNIa}=\mathcal{M}^{\dag}\cdot\mathcal{C}_{SNIa}^{-1}\cdot\mathcal{M}+\mathcal{A}_{SNIa}+\mathcal{B}_{SNIa} ,
\end{eqnarray}
where $ \mathcal{M}\equiv \mu_{obs}(z)-25-5\log_{10}[H_0d_L(z;\{ \mathcal{P}\})/c] $, and
\begin{eqnarray}
\mathcal{A}&\equiv&-\frac{\left[\sum_{i,j}\mathcal{M}(z_i;\{\mathcal{P}\})\mathcal{C}^{-1}_{SNIa}(z_i,z_j)-\ln 10/5\right]^2}{\sum_{i,j}\mathcal{C}^{-1}_{SNIa}(z_i,z_j)}  ,            \\
\mathcal{B}&\equiv&-2\ln\left( \frac{\ln 10}{5}\sqrt{\frac{2\pi}{\sum_{i,j}\mathcal{C}^{-1}_{SNIa}(z_i,z_j)}}\right) .
\end{eqnarray}
The observed distance modulus and the relevant covariance matrix can
be found on the website \cite{JLAdata1,JLAdata2}.

\subsection{Hubble constant}
CMB includes mostly physics at the epoch of recombination, and so provides only weak direct constraints about low-redshift quantities through the integrated Sachs-Wolfe effect and CMB lensing. The CMB-inferred constraints on the local expansion rate $H_0$ are model dependent, and this makes the comparison to direct measurements interesting, since any mismatch could be evidence of new physics. Subsequently, we use Hubble constant measurement from Hubble Space Telescope (HST) $(H = 73.8 \pm 2.4 km s^{-1}Mpc^{-1})$ with flat prior  \cite{Riess:2011yx}.

\subsection{Baryon Acoustic Oscillations}
Baryon acoustic oscillations (BAO) are the imprint of oscillations produced in the baryon-photon plasma on the matter power spectrum. It reveals a standard ruler, but it is calibrated to the CMB-determined sound horizon at the end of the decoupling. The characteristic scale of this pattern is 152 Mpc in comoving scale. On this scale, matter fluctuations experience almost their linear regime. BAO criterion is almost sensitive to the dark energy and matter densities. Therefore, adding BAO prior in our analysis enables to reduce some degeneracies on cosmological parameters, especially for $\Omega_ch^2$ and $H_0$.  The BAO data can be applied to measure both the angular diameter distance $D_A(z;\{\mathcal{P}\})$, and the expansion rate of the Universe $H(z;\{\mathcal{P}\})$. The combination of mentioned quantities is defined by \cite{Ade:2015rim}:
\begin{equation}
D_V(z;\{\mathcal{P}\}) = \left[ (1+z)^2 D_A^2 (z;\{\mathcal{P}\}) \frac{cz}{H(z;\{\mathcal{P}\})} \right]^{1/3} ,
\end{equation}
where $D_V (z;\{\mathcal{P}\})$ is volume-distance. The distance ratio in the context of  BAO criterion is defined by:
\begin{eqnarray}
d_{BAO}(z;\{\mathcal{P}\})\equiv \frac{r_{s}(z;\{\mathcal{P}\})}{D_V(z;\{\mathcal{P}\})}.
\end{eqnarray}
Hence, $r_{s}(z;\{\mathcal{P}\})$ is the comoving sound horizon.  BAO observations contain 6 measurements from redshift interval, $z\in [0.1,0.7]$. In this paper, we use 6 measurements of BAO indicator including Sloan Digital Sky Survey (SDSS) data release 7 (DR7) \cite{Padmanabhan:2012hf},  SDSS-III Baryon Oscillation Spectroscopic Survey (BOSS) \cite{Anderson:2012sa}, WiggleZ survey \cite{Blake:2012pj} and 6dFGS survey \cite{Beutler:2011hx}. In Table \ref{tbl:baodata} we report the observed values for BAO. \\
\begin{table}[ht]
\centering
\begin{tabular}{llcccc}
\hline
\hline {\rm Redshift} & {\rm Data Set} & $d_{obs}$ & {\rm Ref.}\\ \hline
0.10   &  6dFGS     & $0.336\pm0.015$       &  \cite{Beutler:2011hx} \\
0.35 &  SDSS-DR7-rec    & $0.113\pm0.002$& \cite{Padmanabhan:2012hf} \\
0.57 &  SDSS-DR9-rec    & $0.073\pm0.001$ & \cite{Anderson:2012sa} \\
0.44 &  WiggleZ & $0.0916\pm0.0071$     &  \cite{Blake:2012pj} \\
0.60 &  WiggleZ & $0.0726\pm0.0034$     &  \cite{Blake:2012pj} \\
0.73 &  WiggleZ & $0.0592\pm0.0032$     &  \cite{Blake:2012pj} \\
\hline
\hline
\end{tabular}
\caption{ \label{tbl:baodata} Observed data for BAO \cite{hinshaw12}. }
\end{table}
The $d_{obs}(z)$ is reported in Table \ref{tbl:baodata}.
The  $\chi^2_{BAO}$ is defined by
\begin{eqnarray}
\chi^2_{BAO}\equiv \Delta d^{\dag}\cdot {\mathcal
C}^{-1}_{BAO}\cdot \Delta d\,.
\end{eqnarray}
here $\Delta d(z;\{\mathcal{P}\})\equiv
d_{obs}(z)-d_{BAO}(z;\{\mathcal{P}\})$, and  inverse of covariance matrix, ${\mathcal C}^{-1}_{BAO}$, is \cite{Hinshaw:2012aka}:
\begin{eqnarray}\label{covbao}
{\mathcal C}^{-1}_{BAO} = \left(\begin{array}{rrrrrr}
4444.4 & 0 & 0 & 0 & 0 & 0 \\
0 & 34.602 & 0 & 0 & 0 & 0 \\
0 & 0 & 20.661157 & 0 & 0 & 0 \\
0 & 0 & 0 & 24532.1  & -25137.7 & 12099.1 \\
0 & 0 & 0 & -25137.7 & 134598.4 & -64783.9 \\
0 & 0 & 0 & 12099.1 & -64783.9 & 128837.6 \\
\end{array}
\right).
\end{eqnarray}

\subsection{CMB observations}

The CMB data alone is not able to put constraint on free parameters
of dark energy models well. Because the main effects of dark energy
constraint in the CMB anisotropy spectrum come from an angular
diameter distance to the decoupling epoch $z \simeq 1100$ and the
late integrated Sachs-Wolfe (ISW) effect. The late ISW effect cannot
be accurately measured currently, therefore, only important
information for constraint dark energy in the CMB data actually
comes from the angular diameter distance to the last scattering
surface However, if we consider clustering of DDE, one can expect to
get non-trivial behavior and it needs to modify Boltzmann code to
compute evolution of perturbations.  In this work we focus on models
which mainly affect the expansion history. To include all the
aspects of models in cosmic evolution we use full CMB data from
$Planck$ mission. In this part we use observed CMB power spectrum.
The chi-square for this observation is:
\begin{eqnarray}
\chi^2_{CMB-power}=\Delta C^{\dag}\cdot \mathcal {M}^{-1}\cdot \Delta C ,
\end{eqnarray}
here $\Delta C_{\ell}\equiv C_{\ell}^{obs}-C_{\ell}(\{\mathcal{P}\})$ and $\mathcal {M}$ is covariance matrix for CMB power spectrum.  We also utilize CMB lensing from SMICA pipeline of $Planck$ 2015. To compute CMB power spectrum for our model, we modify Boltzmann code CAMB \cite{Lewis:1999bs}.
\subsection{$f\sigma_8$ observations}
Using the amplitude of over-density at the comoving $8 h^{-1} \rm Mpc$ scale and in order to obtain constraints from RSD measurements, The following quantity introduced as:
\begin{equation}
y(z) \equiv f(z) \sigma_8(z) .
\end{equation}
Throughout this paper we use most recent observational data sets based on SDSS-III, SDSS-II, DR7, VIPERS, RSD projects provided for $f(a)\sigma_8(a)$ to check the performance of DDE models (see  \cite{Geng:2017apd,data-1} which collected some points for $f\sigma_8$ at different redshifts and references therein).

Finally the total value of chi-square reads as:
\begin{equation}
\chi^2 = \chi^2_{SNIa}+ \chi^2_{BAO} +\chi^2_{HST}+\chi^2_{CMB}+\chi^2_{f\sigma_8} .
\end{equation}
We utilize publicly available cosmological  Markov Chain Monte Carlo code CosmoMC  \cite{Lewis:2002ah}  to find a global fitting on the cosmological parameters in models. For CMB part we combine CAMB with CosmoMC. To take into account the rest of observational quantities, we write our code in {\bf  python}.

\begin{table}[t]
\begin{center}
\centering
\begin{tabular}{|l||c|c|c|c|}
\hline
 \hline Parameter  &PL & CPL & FSL  \\
\hline
$\Omega_b h^2 $ &$0.02217\pm 0.00015$& $0.02218\pm 0.00016$&$0.02214\pm 0.00015 $ \\
\hline
$\Omega_{\rm c}h^2$ &$0.1195\pm 0.0013$ &$0.1193^{+0.0014}_{-0.0017}$&$0.1190^{+0.0013}_{-0.0016}$ \\
\hline
$\Omega_{\rm DE}$  & $0.6862\pm 0.0078$&$0.688^{+0.010}_{-0.0079}$&$0.6893\pm 0.0078$  \\
\hline
$\alpha$&$ 0.1013\pm 0.0031$&$-$&$-$\\
\hline
$w_0 $  & $-1.3799^{+0.0036}_{-0.0028} $&$-1.08045^{+0.00041}_{-0.00062}$&$-0.9994\pm 0.0076$\\
\hline
$w_1 $ &$-$&$-0.12190^{+0.00050}_{-0.00030}$&$-0.0082^{+0.0044}_{-0.0051}$ \\
\hline
$H_0 $ & $67.36\pm 0.56$&$ 67.48^{+0.71}_{-0.57}$&$67.56\pm 0.54$   \\
\hline
$\Theta_{\rm MC} $ &$1.04090^{+0.00072}_{-0.00076}  $&$1.04094^{+0.00085}_{-0.00089}$&$1.04099^{+0.00068}_{-0.00073}$   \\
\hline
$\tau $ & $0.0828^{+0.0021}_{-0.0030}$&$0.0832^{+0.0028}_{-0.0034}$&$0.0837^{+0.0036}_{-0.0042}$\\
\hline
$n_s $ & $0.9665^{+0.0079}_{-0.0076}  $&$0.9641^{+0.0068}_{-0.0070}$&$0.9659^{+0.0074}_{-0.0078}$  \\
\hline
$\ln \left( 10^{10} A_s \right) $ & $3.1006\pm 0.0038$&$3.0992^{+0.0065}_{-0.0070}$&$3.1009\pm 0.0041$ \\
\hline
\hline
\end{tabular}
\caption{\label{tbl:results-cmbbsh} The best fit values at $1\sigma$ confidence interval for free parameters of models based on joint analysis of SNIa+BAO+HST+$Planck$ TT+ LSS. }
\end{center}
\end{table}

\section{Results and Discussion}
In this section we will present the results of observational constraints for the CPL, FSL and PL as DDE models with the following  observational data sets:  740 SNIa from JLA catalogue, the Hubble parameter at present time form HST, 6 data points for BAO, full sky CMB temperature power spectrum from $Planck$ 2015, $f\sigma_8$ observations  by SDSS-II, SDSS-III, DR7, VIPERS and RSD projects. Then, we discuss our  result  for the three DDE models .\\

\begin{figure}[t]
\centering
\includegraphics[width=1\textwidth]{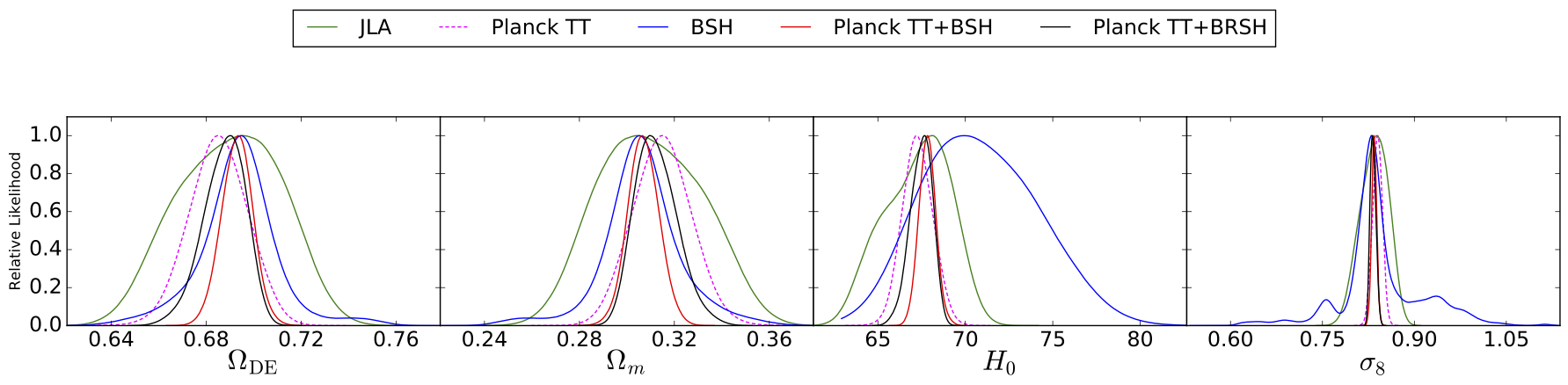}
\caption{Marginalized posterior function for $\Omega_{\rm DE}$, $\Omega_m$, $H_0$ and $\sigma_8$ for CPL model. Solid green line corresponds to observational constraint by SNIa using JLA catalogue. Dashed pink line indicates constraint by $Planck$ TT data set. Blue line is devoted to joint analysis BAO+SNIa+HST (BSH). Red line is for joint analysis of $Planck$ TT+BSH. Combination of all observations is illustrated by thick solid black line. }
\label{1D_CPL}
\end{figure}

\begin{figure}
\centering
\includegraphics[width=0.95\textwidth]{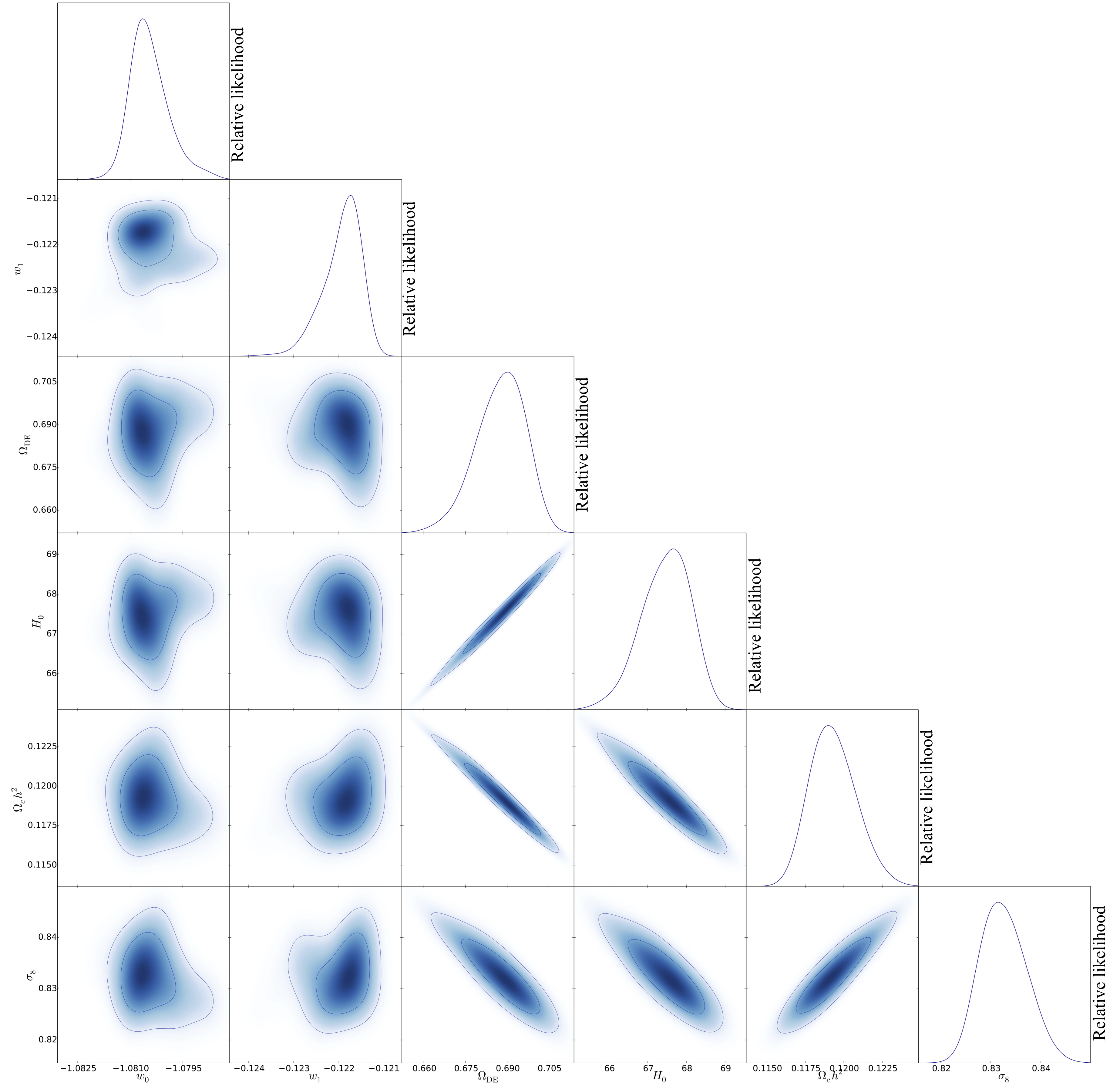}
\caption{Measurement of various free parameters of CPL model from combination of SNIa+BAO+HST+$Planck$ TT+LSS  observation. The 2-D regions with $1\sigma$ and $2\sigma$ level of confidences and corresponding 1-D marginalized posterior function are indicated in this figure.}
\label{CPLcontour}
\end{figure}

{\bf CPL model:}\\
Combining all observational data sets namely, SNIa+BAO+HST+$Planck$ TT+LSS leads to $w_0=-1.08045^{+0.00041}_{-0.00062}$ and $w_1=-0.12190^{+0.00050}_{-0.00030}$ at $1\sigma$ confidence interval. Table \ref{tbl:results-cmbbsh} present the other free cosmological parameters of the model.  Figs. \ref{1D_CPL} and \ref{CPLcontour}
illustrate marginalized posterior probability  distribution and the contours for the various parameters, receptively.

\begin{figure}[t]
\centering
\includegraphics[width=1\textwidth]{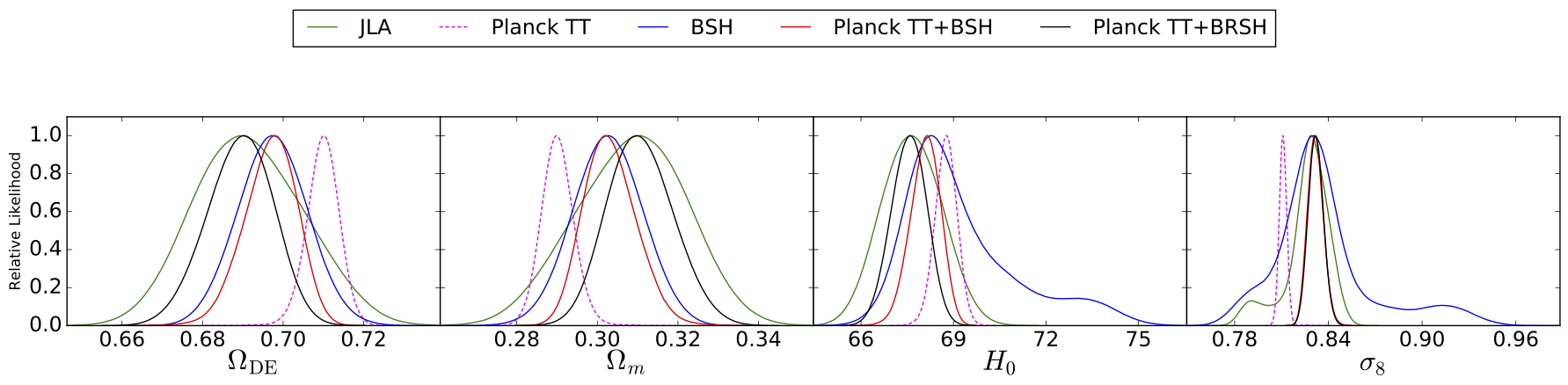}
\caption{Marginalized posterior function for $\Omega_{\rm DE}$, $\Omega_m$, $H_0$ and $\sigma_8$ for FSL model. Solid green line corresponds to observational constraint by SNIa using JLA catalogue. Dashed pink line indicates constraint by $Planck$ TT data set.  Blue line is devoted to joint analysis BAO+SNIa+HST (BSH).Red line is for joint analysis of $Planck$ TT+BSH. Combination of all observations is illustrated by thick solid black line. }
\label{1D_FSL}
\end{figure}

\begin{figure}[h]
\centering
\includegraphics[width=0.95\textwidth]{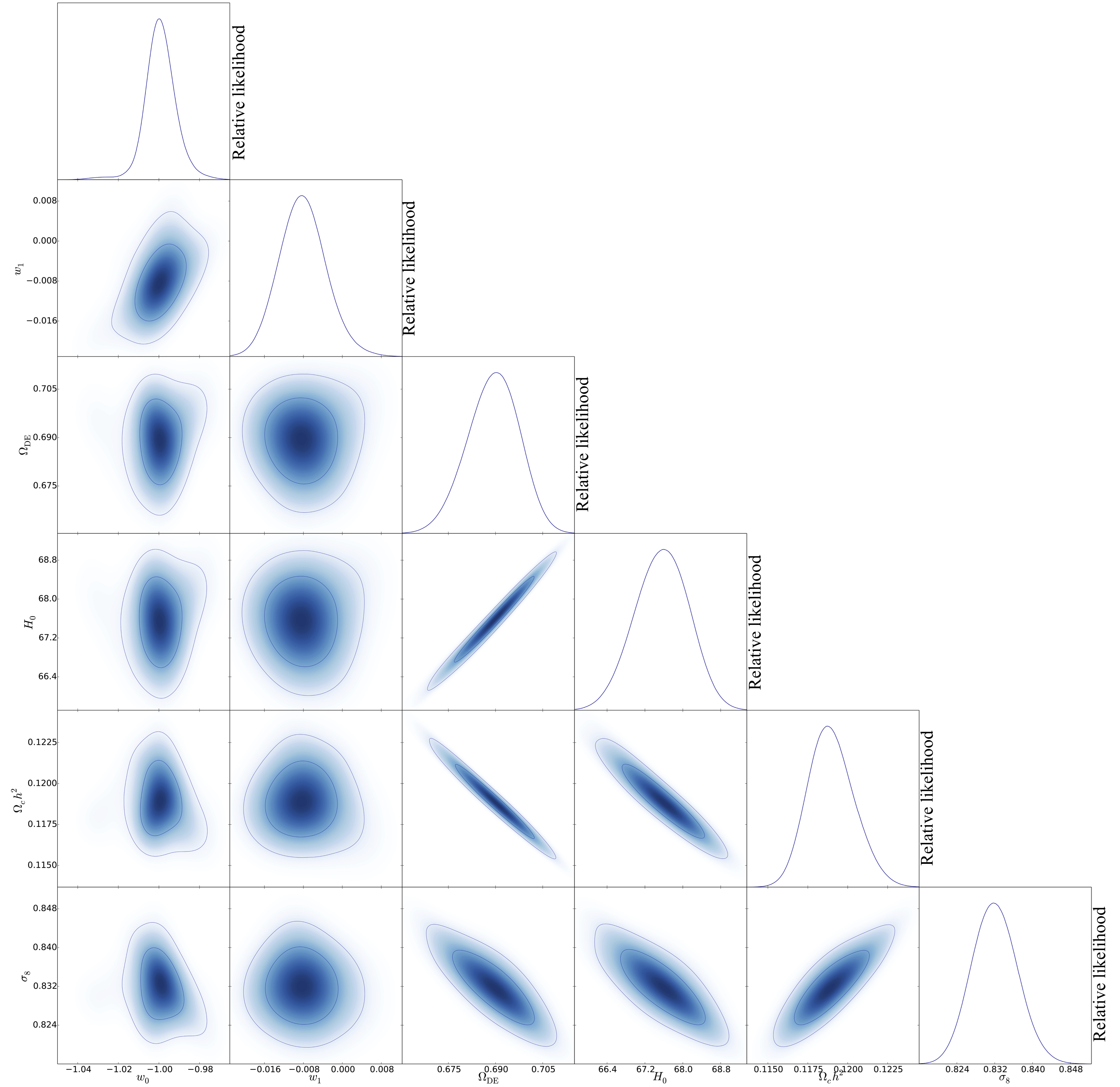}
\caption{Measurement of various free parameters of FSL model from combination of SNIa+BAO+HST+$Planck$ TT+LSS  observation. The 2-D regions with $1\sigma$ and $2\sigma$ level of confidences and corresponding 1-D marginalized posterior function are indicated in this figure.}
\label{FSLcontour}
\end{figure}

{\bf FSL model:}\\
The best fit parameter of this model for the joint analysis of  SNIa+BAO+HST+$Planck$ TT+LSS observation are: $w_0=-0.9994\pm 0.0076$ and $w_1=-0.0082^{+0.0044}_{-0.0051}$ at $1\sigma$ confidence interval. The rest of the parameters are given in table \ref{tbl:results-cmbbsh}.  Figs. \ref{1D_FSL} and \ref{FSLcontour} indicate the marginalized posterior probability  distribution and contours for  the various free parameters, respectively.



\begin{figure}[t]
\centering
\includegraphics[width=1\textwidth]{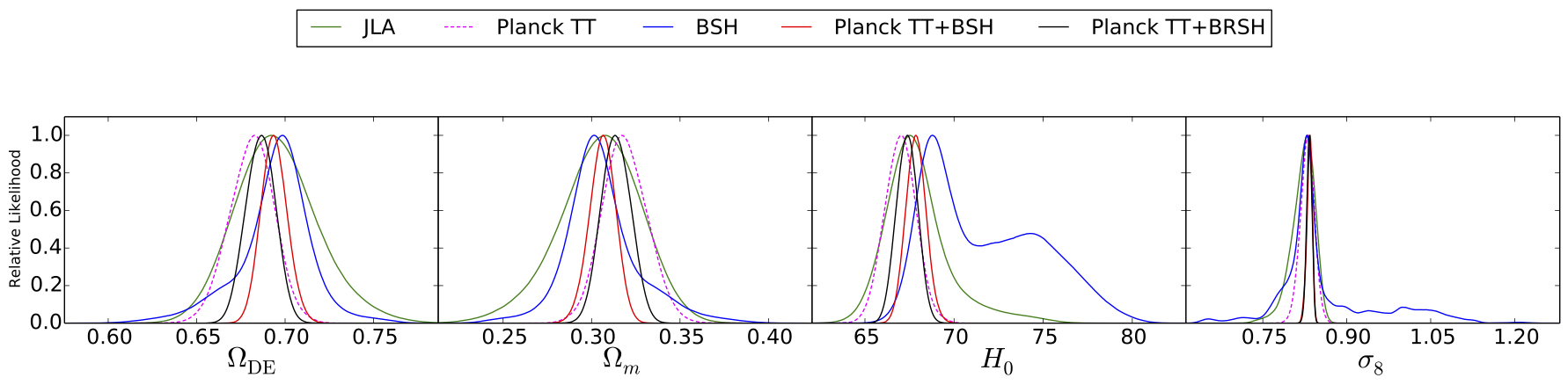}
\caption{Marginalized posterior function for $\Omega_{\rm DE}$, $\Omega_m$, $H_0$ and $\sigma_8$ for PL model. Solid green line corresponds to observational constraint by SNIa using JLA catalogue. Dashed pink line indicates constraint by Planck TT data set. Blue line is devoted to joint analysis BAO+SNIa+HST (BSH). Red line is for joint analysis of $Planck$ TT+BSH. Combination of all observations is illustrated by thick solid black line. }
\label{1D_PL}
\end{figure}

\begin{figure}[h]
\centering
\includegraphics[width=0.95\textwidth]{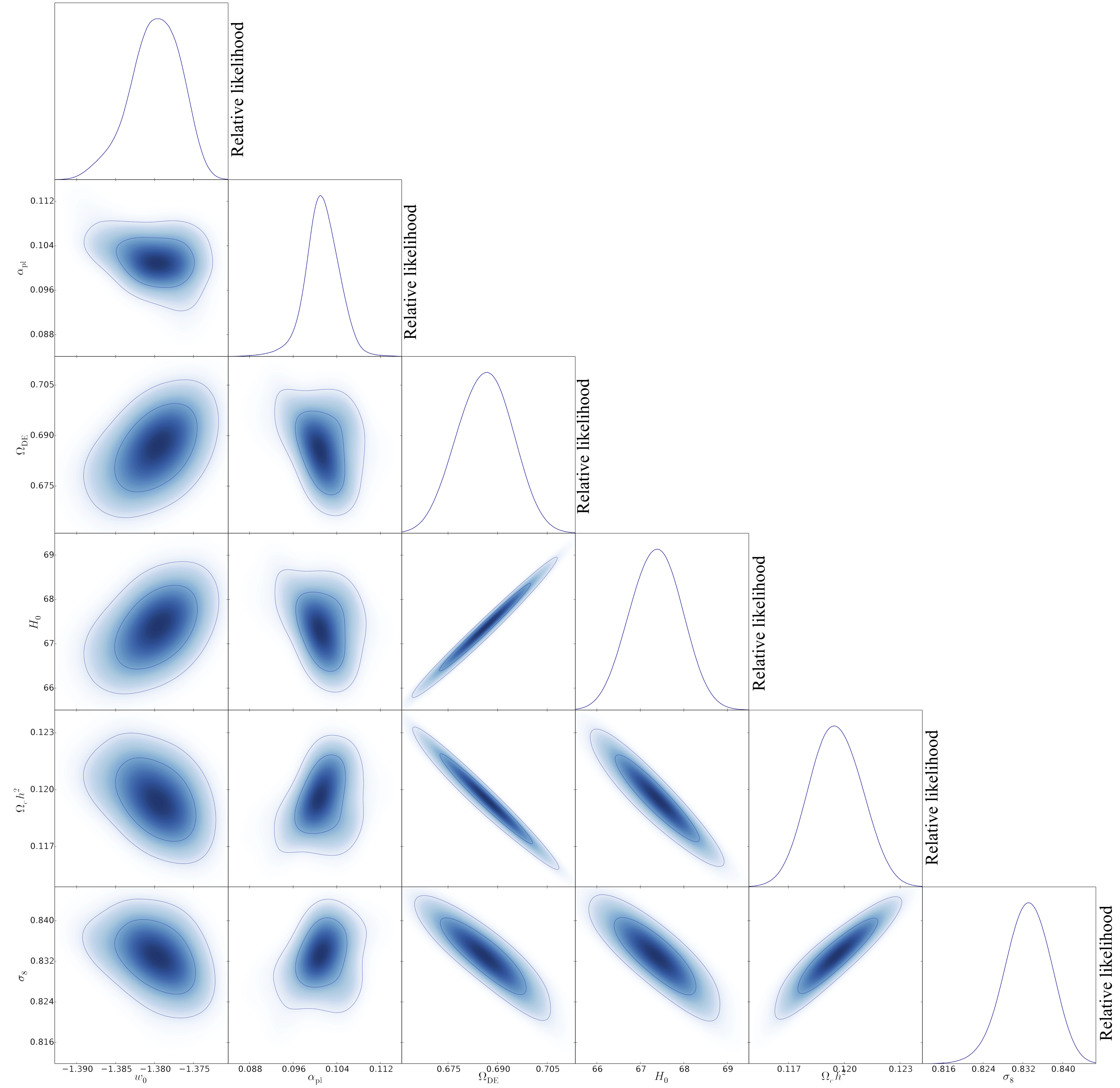}
\caption{Measurement of various free parameters of PL model from combination of SNIa+BAO+HST+$Planck$ TT+LSS  observation. The 2-D regions with $1\sigma$ and $2\sigma$ level of confidences and corresponding 1-D marginalized posterior function are indicated in this figure.}
\label{PLcontour}
\end{figure}

\begin{figure}[t]
\centering
\includegraphics[width=.43\textwidth,origin=c,angle=0]{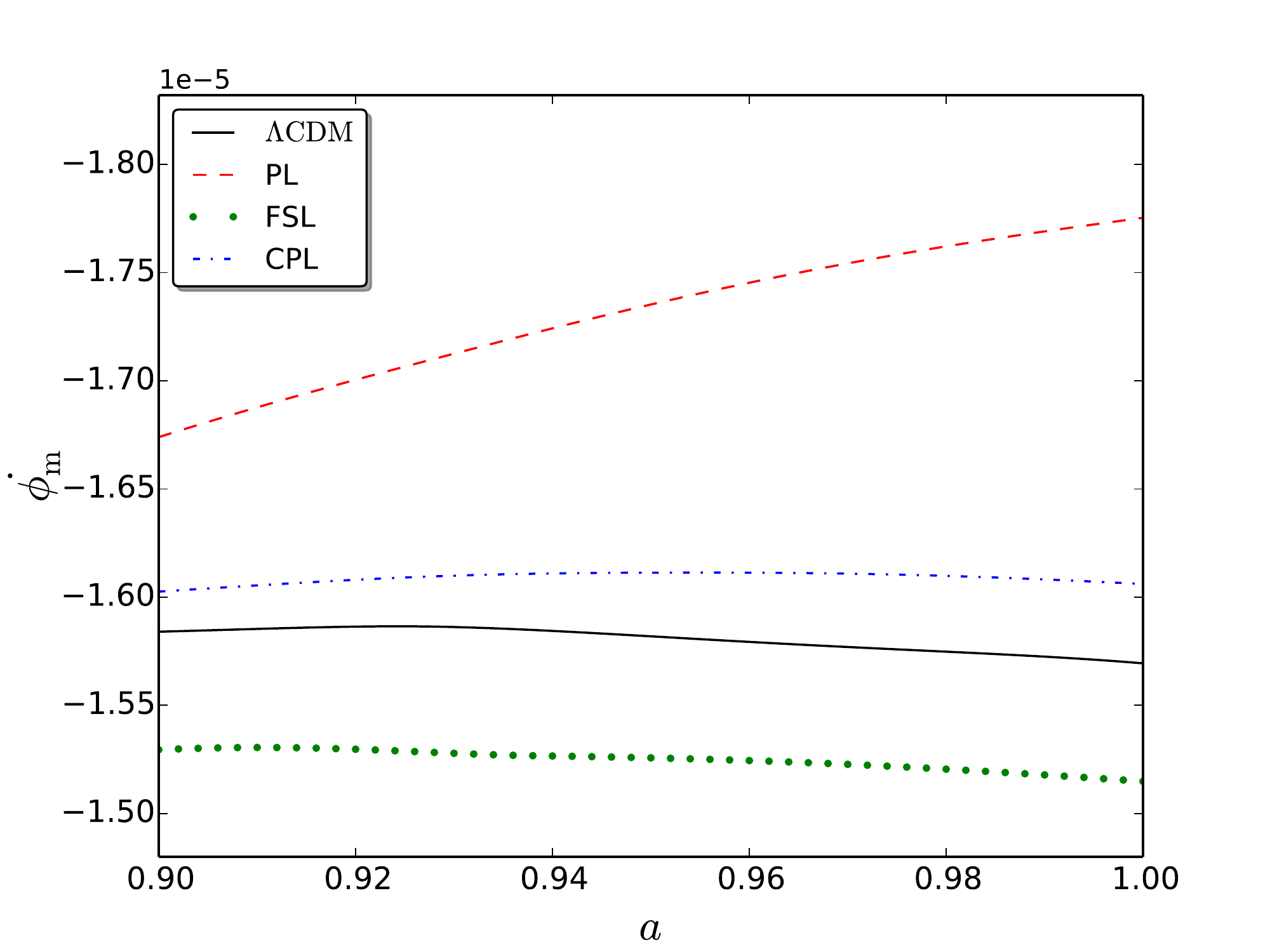}
\includegraphics[width=.5\textwidth,origin=c,angle=0]{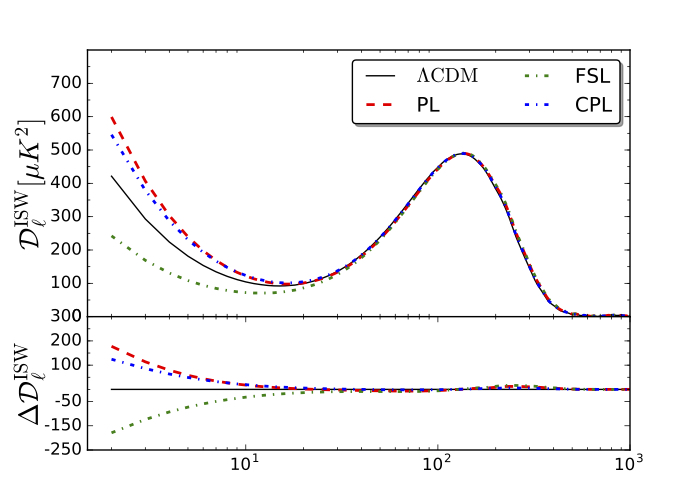}

\caption{\label{fig:Dpotential}Left panel corresponds to evolution of time derivative of matter potential for various dark energy models versus scale factor in the linear regime. Right panel shows pure ISW effect as a function of $\ell$. The lower right panel illustrates difference of pure ISW power spectrum with $\lambda$CDM to make more obvious their deviations. We take best fit values for free parameters determined by combining SNIa+$Planck$ TT+ BAO+ HST+LSS data sets.  }
\end{figure}

{\bf PL model:}\\
The parameter of  the  power law model using the joint SNIa+BAO+HST+$Planck$ TT+LSS  are  $w_0=-1.3799^{+0.0036}_{-0.0028}$ and $\alpha=0.1013\pm 0.0031$ at $1\sigma$ confidence interval. Standard value for the standard cosmic parameters are given in  table \ref{tbl:results-cmbbsh}.   Figs. \ref{1D_PL} and \ref{PLcontour} depict the marginalized posterior function and contours for various free parameters, respectively. The tension in  $\sigma_8$ and $ H_{0} $ as measured  by {\it Planck} TT and late time observations, is almost alleviated in PL model.

\begin{figure}[t]
\centering
\includegraphics[width=.42\textwidth,origin=c,angle=0]{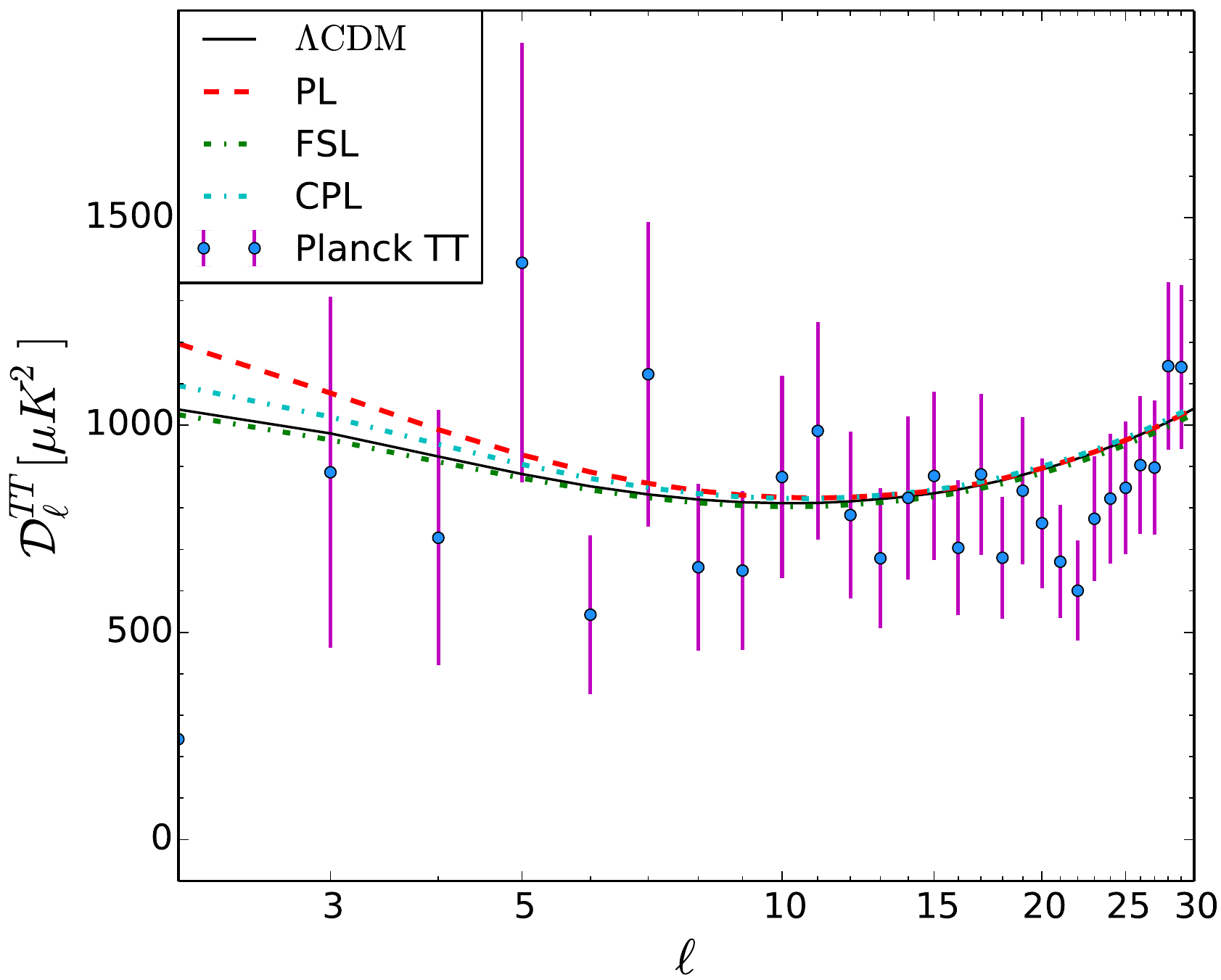}
\includegraphics[width=.49\textwidth,origin=c,angle=0]{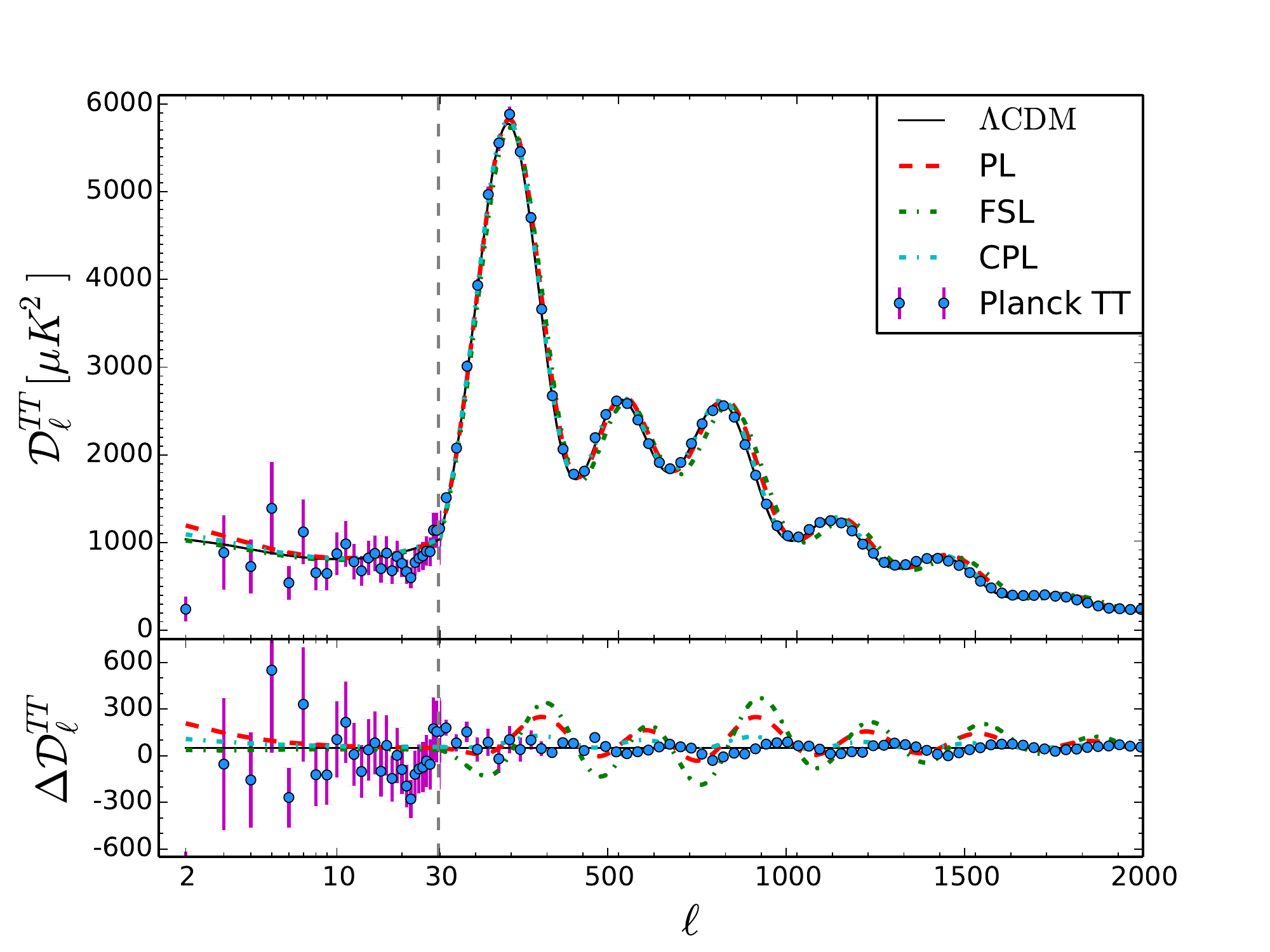}

\caption{\label{cl} Left panel corresponds to temperature power spectrum of three DDE models at low $\ell$. Right panel indicates the TT power spectrum of CMB. Residuals with respect to  $\Lambda$CDM model are shown in the lower panel. The error bars show at $1\sigma$  confidence interval. Best fit values for free parameters have been achieved by joint analysis of JLA+BAO+HST+CMB+LSS.}
\end{figure}

It is also interesting to investigate the  dynamical behavior of the
matter potential which has the prominent role in producing the late
ISW on the CMB anisotropy.  Using the equation of state for the
models compared to cosmological constant, we find that the
associated matter potentials at linear regime for CPL and FSL follow
closely the $\Lambda$CDM model. However PL model equation of state
crossing $\Lambda$CDM and cause to achieve higher value for
$\dot{\phi}_m$ all the time. The left panel of Fig.
\ref{fig:Dpotential} indicates $\dot{\phi}_m$ as a function of scale
factor for the best fit values given in table
\ref{tbl:results-cmbbsh} by joint analysis. The right panel of Fig.
\ref{fig:Dpotential} compute the ISW contribution to the CMB power
spectrum for three DDE models. The lower part of right panel of Fig.
\ref{fig:Dpotential}  right panel shows deviation of late ISW
contribution in the presence of CPL, FSL and PL models with respect
to $\Lambda$CDM model .It demonstrate higher value for PL and lower
value for FSL as expected from variation of potential. The full CMB
temperature power spectrum $\mathcal{D}_{\ell}^{TT}$, with all
source of the anisotropy included, is plotted in Fig. \ref{cl}. The
these DE models are consistent with data and not distinguishable
from $\Lambda$CDM model with the given error bars. The Left panel
Fig. \ref{fig:Dpotential} exhibits highest ISW contribution for PL
compared to other DDE models, although the uncertainty due to cosmic
variance cause that the best model is not measurable.

\begin{figure}[t]
\centering
\includegraphics[width=.48\textwidth,origin=c,angle=0]{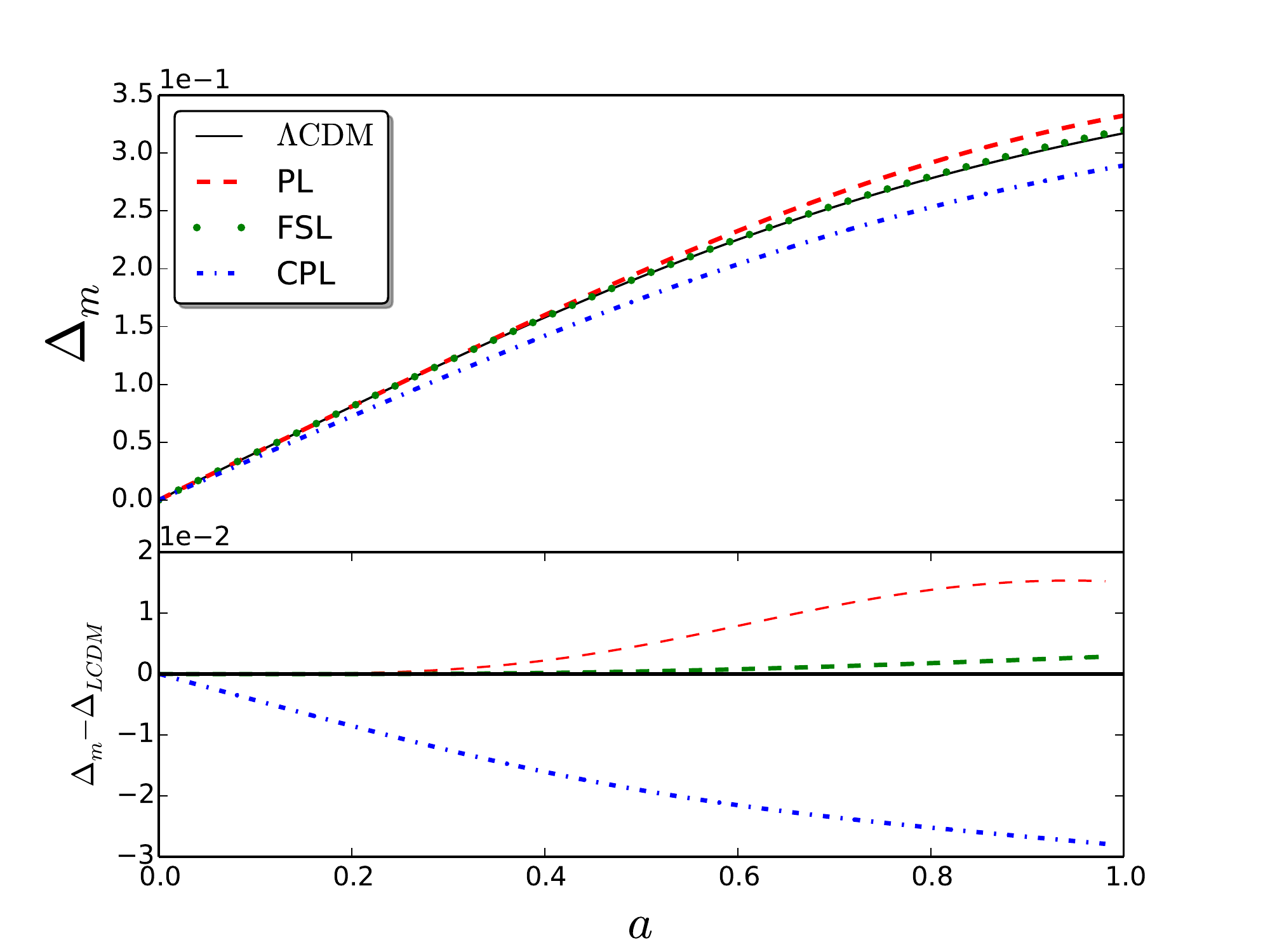}
\includegraphics[width=.48\textwidth,origin=c,angle=0]{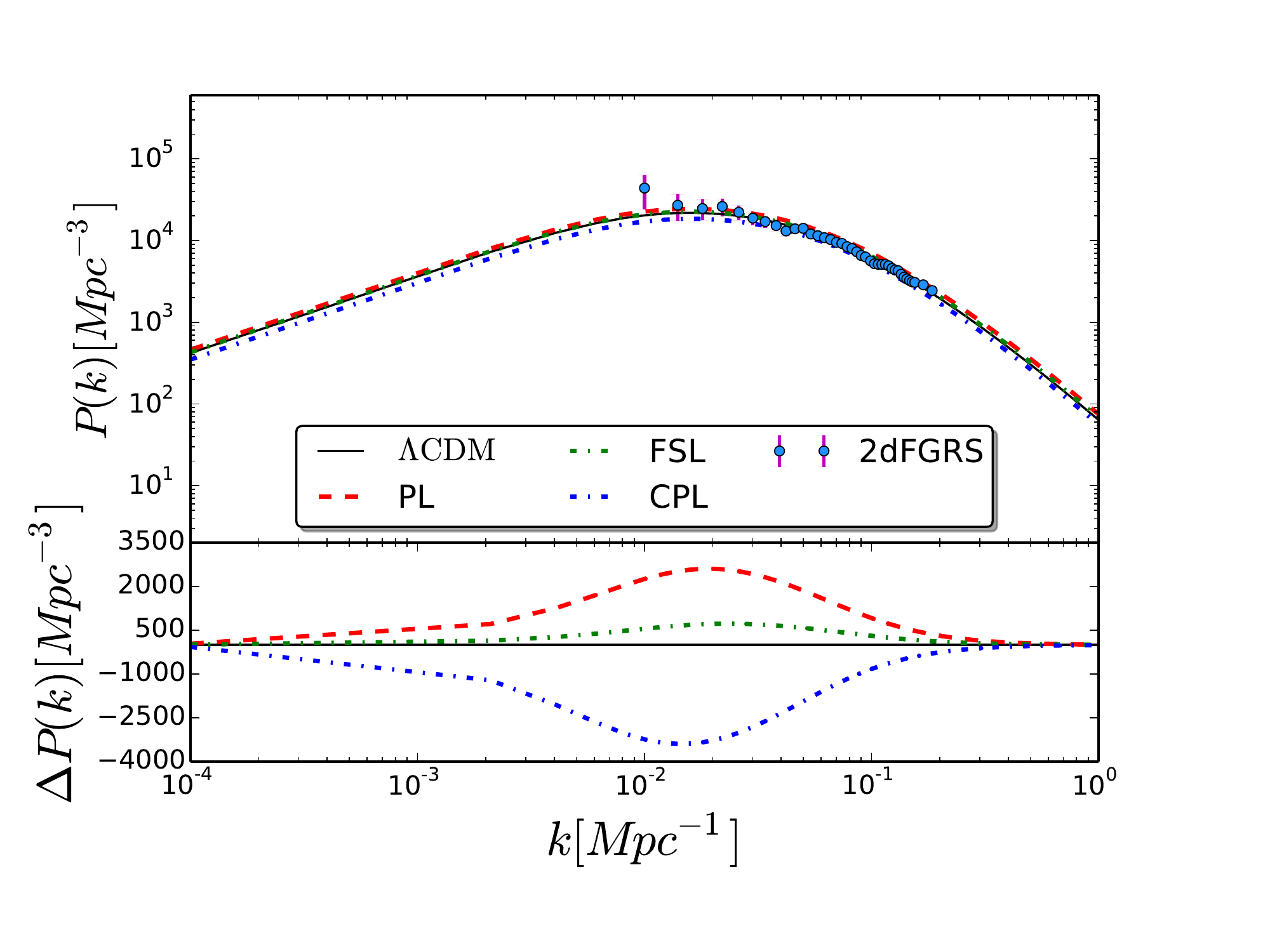}

\caption{\label{fig:matt}Left panel shows evolution of density contrast versus scale factor. The lower part of this panel represents difference between $\Delta_m$ and that of computed for $\Lambda$CDM model. The right panel corresponds to matter power spectrum of DDE models and for $\Lambda$CDM. Here to make more sense we show the 2dFGR data for matter power spectrum indicated by filled circle symbols. We compute the difference between $P(k)$ of DDE model an $\Lambda$CDM to make more sense in the lower part of this panel. Best fit values for free parameters have been determined by combining SNIa+ BAO+ HST+$Planck$ TT+LSS data sets.}
\end{figure}

Fig. \ref{fig:matt} shows the evolution of the matter density
contrast $ \Delta_{m} $ defined by Eq. (\ref{eq:Deltaobs}) and the
matter power spectrum computed for three DDE models. As illustrated
in the left panel of Fig. \ref{fig:matt}, we find an enhancement in
the matter growth for the PL model. This can be justified by
comparing the values of $\Omega_{\rm DE}$ of the PL and $\Lambda$CDM
models and by looking at the equation of state. The value of
$\Omega_{\rm DE}$ for PL model is less than that  of computed for
$\Lambda$CDM model. The $\bar{w}_{\rm PL}$ behaves effectively
similar to matter's equation of state. On the other hand, there is
suppression in the matter growth for the CPL model with the best fit
parameters compared to $\Lambda$CDM model.  For FSL, $ \Delta_{m} $
is almost similar to the concordance model. The lower left panel of
Fig. \ref{fig:matt} represents the relative difference of observable
density contrast to $\Lambda$CDM. The right panel of Fig.
\ref{fig:matt} indicates the matter power spectrum for various DDE
models . We observe an excess of matter power spectrum for the  PL
model compared to the  $\Lambda$CDM model especially at intermediate
scales, namely $10^{-3}<k<2\times 10^{-1}$, with lower amount for
the FSL and CPL models. This results is expected given the crossing
behavior of the PL model. At  early epoch, dark energy in PL
contributes as cold dark matter leading to get more enhancement for
structure formation.  Also,the PL model exhibit strong clustering at
lower value of scale factor and decrease with scale factor until
present time. Such behavior may alter the cause for some of
observational discrepancies such as missing satellite, which we
postponed for next work trough N-body simulation of DDE models.

\begin{figure}[t]
\centering
\includegraphics[width=.5\textwidth,origin=c,angle=0]{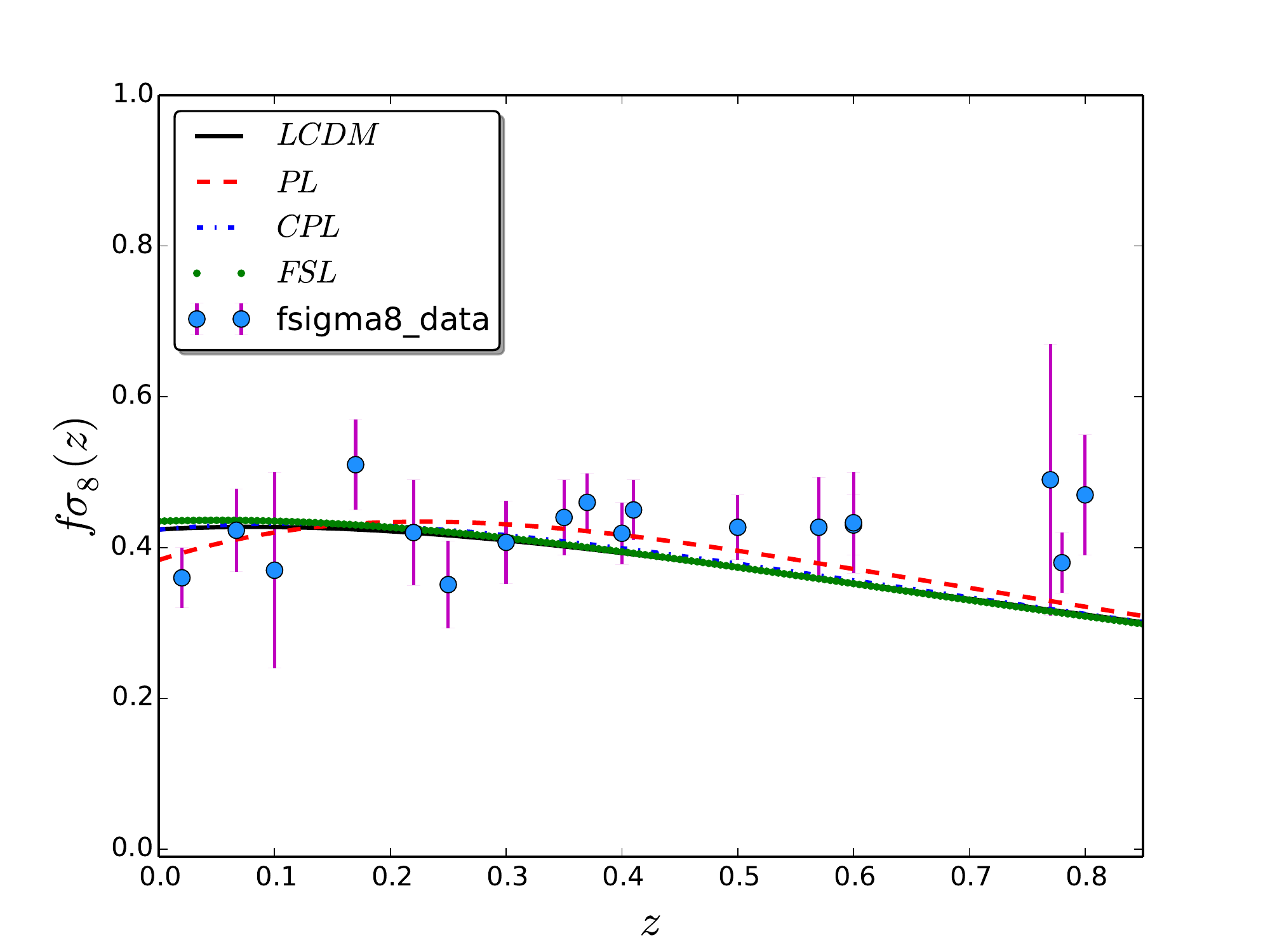}
\caption{\label{fig:fsigma8}The model independent parameter for linear growth rate of cosmic structure in the present of dynamical dark energy models. Theoretical prediction indicated in this plot used the best fit values for free parameters achieved by joint analysis of JLA+BAO+HST+CMB+LSS.}
\end{figure}

We turn to examine the $f\sigma_8(z)$ for three DDE models explained in this paper.  As shown in Fig. \ref{fig:fsigma8}, PL model has more consistency compared with other models. Our results demonstrates that, future observations with more precise accuracy enable us to discriminate PL model from $\Lambda$CDM-like models.

Now, we deal with examining relevant quantity coming from perturbation theory for our DDE models and we will compare them with $\Lambda$CDM model. As discussed in section  \ref{DDE clustering}, and plotted \ref{fig:DEC}, the density contrast of dark energy component behaves differently as function of scale factor for various DDE models. In particular, the $ \Delta_{DE} $ for PL is positive at all time with maximum of  $ \Delta_{PL}\simeq10^{-3} $, while for other two models density contrast of dark energy  has several crossing $ \Delta_{DE}=0 $ at various time depending on the model.the sound speed show how fast  pressure perturbations propagate through the fluid. Higher values of $c_s^2$ leads to a decreased of clustering of dark energy  for given initial perturbations. The plot of   Fig. \ref{fig:DEC} also show the effect of sound speed on clustering of dark energy with different line style for DDE models.
The CPL and FSL models show small clustering and exhibit dark energy void   with deep valleys at the early universe.we also find that the FSL  and CPL have a valley at early universe whose depth depends on sound speed. Reducing the sound speed increase the depth of valley and shifted it  to late time.\\

The number density of peaks if we consider Gaussian random field for CMB in the presence of DDE models are indicated in Fig. \ref{fig:ND}.  There are potential capability for discrimination of various DDE models for number density of peaks as a function of threshold at far from mean threshold. At $ \nu=-2 $ there is most separation between models, hence empty and dense regions are beneficial for discrimination of DDE models.

 In PL dynamical dark energy model, we have $ \Omega_{\rm PL}(z_{lss})\approx 10^{-1}$ while  $ \Omega_{\rm{CPL},\rm{FSL}}(z_{lss})\approx 10^{-12}$. Therefore, one can not ignore the contribution of PL dark energy model at early era leading to some beneficial differences for upcoming data.
Number of clusters expected by Euclid satellite enables to indicate detectable signal if EDE to be existed \cite{DEC-1}.
A robust indicator for searching the footprint of EDE is that: galaxy power spectrum amplitude at spatial scale greater than sound horizon is affected by dark energy clustering causing an enhancement which is sensitive to redshift evolution of net dark energy density (i.e. the equation of state) \cite{DEC-4}.
According \ref{1D_CPL}, \ref{1D_FSL}, \ref{1D_PL} DDE models can improve $ H_{0} $ tension, which Relative Likelihood of PL  for Planck and SNIa completely cover each other and disappear the tension. Although CPL improved tension but there is separation between peak of two data set. FSL is not suitable model for $ H_{0} $ tension.

\begin{table}[t]
\begin{center}
\centering
\begin{tabular}{|c| c cc c|}
\hline \hline
$\rm{Model}$ & $\chi^2_{min}$ & $\Delta\rm{AIC}$ & $\Delta\rm {BIC}$& \\
\hline
$\Lambda$CDM  &1629.69& 0& 0&\\
\hline
$\rm{CPL}$  & 1631.57& 3.88  & 18.077&\\
\hline
$\rm{FSL}$  &1629.87 &2.18 &17.007& \\
\hline
$\rm{PL}$  & 1638.73& 11.40&25.597& \\
\hline
\end{tabular}
\caption{\label{tb:aic}The minimum value of $\chi^2$, $\Delta$AIC and $\Delta$BIC criteria for
our models and $\Lambda$CDM when we use SNIa+ BAO+ HST+$Planck$ TT+LSS data sets. }
\end{center}
\end{table}

There are standard technique for quantity comparison of  different
models with different parameter describing the same data sets. We
will use some of these technique to compare the DDE models with each
other and with $\Lambda$CDM model as a reference model. We need to
other quantity in addition to determine $ \chi^{2} $.  It turns out
that  more degrees of freedom usually lead to better fit of the data
with assumed model. However, there should be a penalty for adding to
the  of the model by introducing more parameters.
 In this work, we use $ \chi^{2} $,  AIC~\cite{H.Akaike:1974} and BIC~\cite{G.schwarz:1978} criteria for model comparison.
 AIC is defined by:
\begin{equation}
{\rm AIC}=-2\ln{\mathcal{L}_{\rm{max}}}+2k,
\end{equation}
and BIC~\cite{G.schwarz:1978} is given by:
\begin{equation}
{\rm BIC}=-2\ln{\mathcal{L}_{\rm{max}}}+k\ln{N},
\end{equation}

In the above equations, $k$ is the number of free parameters. Note
that BIC also depends on $N$,  the number of observational data
point carried out for implying observational constraints
\cite{Xu:2016grp}. In practice, we reported these criteria for DDE
models relative to $\Lambda$CDM , i.e $\Delta {\rm
AIC}=\Delta\chi^{2}_{\rm{min}}+2\Delta g$ and $\Delta{\rm
BIC}=\Delta\chi^{2}_{\rm{min}}+\Delta g\ln{N}$. Lower values of
$\Delta$AIC and $\Delta$BIC, mean that the assumed model explain the
data well. Table 4 report the $\Delta$AIC and $\Delta$BIC of the DDE
models of our interest.
 The relevant quantities for mentioned purpose for our three models accompanying $\Lambda$CDM are reported in Table \ref{tb:aic}. Our results elucidate that our three dark energy models are supported by observations. We find that all considered models in this paper are worse than $\Lambda$CDM model but still are good models. Our results demonstrate that CPL and FSL are supported by observations with value of $\Delta$AIC$<10$. Note that with  $\Lambda$CDM being reference model, the observation do not supported PL model strongly. However, the  $\Lambda$CDM still gives a better description of the data.

\section{Conclusion}
In this paper we consider three valuable dynamical dark energy models, namely, CPL, FSL and PL. We use most recent observational catalogues in order to confine the value of model free parameters. We implement joint light-curve analysis (JLA) for SNIa, baryon acoustic oscillation (BAO) from various surveys, Hubble parameter from HST-key project, {\it Planck} TT power spectrum and $f\sigma_8$ for large scale structure (LSS) observations. The joint analysis of JLA+BAO+HST+CMB+LSS, shows that $\Omega_{\rm DE}=0.6862\pm0.0078$, $\alpha=0.1013\pm0.0031$ and $w_0=-1.3799^{+0.0036}_{-0.0028}$ for Power-Law dynamical dark energy (DDE) model at $1\sigma$ confidence interval. The same observational constraints one optimal variance error for CPL model leads to $\Omega_{\rm DE}=0.6880^{+0.0100}_{-0.0079}$, $w_0=-1.08045^{+0.00041}_{-0.00062}$ and $w_1=-0.12190^{+0.00050}_{-0.00030}$. While for FSL model we find   $\Omega_{\rm DE}=0.6893\pm0.0078$, $w_0=-0.9994\pm0.0076$ and $w_1=-0.0082^{+0.0044}_{-0.0051}$ at $1\sigma$ confidence limit.  The tension in σ8 and H0 as measured by Planck TT and late time observations, is almost alleviated in PL model.

Since the variation of equation of state for PL is more than all DDE models considered in this paper, also due to more contribution of PL on cold matter at the early universe, namely  $ \Omega_{\rm PL}(z_{lss})\approx 10^{-1}$ while $ \Omega_{\rm{CPL},\rm{FSL}}(z_{lss})\approx 10^{-12}$. Therefore, one can not ignore the contribution of PL dark energy model at early era leading to some beneficial differences for upcoming data. Structure formation gives us opportunity  to discriminate between dynamical dark energy models. We find that growth of matter density in PL model is higher than other DDE models. For CPL, a considerable suppression for structure formation is achieved.  The nature of dark energy can be investigated not only by equation of state but also through clustering and sound speed. We also examine the clustering of DDE model by modifying the relevant perturbation equations. The models show imaginary sound speed  just PL exhibits positive value during period $0<a<0.4$ which cause instability in dark energy, so  we consider $ c_{s}^{2}=0.1 $, $ c_{s}^{2}=0.01 $ and $ c_{s}^{2}=0.001 $ at large scale $ k=0.01 Mpc$.
We obtained that, PL has positive clustering and grows fast during the early universe due to crossing behavior of equation of state and clod dark matter manner at the early time. The maximum of PL clustering is $ \Delta_{PL}\simeq10^{-3} $ and decreasing by increasing the sound speed. Hence, PL model has potential to produce more voids rather than other models \cite{DEC-9} which is left for next work through the N-body simulation. The CPL and FSL exhibit small clustering at early time and density contrast of them have several crossing $ \Delta_{DE} =0 $ at various times. The CPL and FSL exhibit void of dark energy with deep and sharp valleys around $ \Delta_{DE}\simeq -10^{-11} $ which depth of valleys are sensitive to sound speed and shifted to early time by increasing sound speed.\\

A geometrical measures namely the abundance of local maxima as a function of threshold for three DDE models elucidate that at far from mean threshold, it is potentially possible to look for a measure to discriminate different cosmological models.

The contribution of PL and CPL for late ISW are significant comparing to cosmological constant and FSL model.
According Figs: \ref{1D_CPL}, \ref{1D_PL}, \ref{1D_FSL}, tension between HST and CMB for $ H_{0} $ disappears for all models especially for PL model. The $f\sigma_8$ measure has been examined for our DDE models. Accordingly, future observations with more precise accuracy enable us to discriminate PL model from $\Lambda$CDM-like models. Not only, existence of early dark energy with 0.9 \% density at \%99 confidence level but also  detect dark energy is cold rather than canonical \%99 confidenc are possible with cluster survey of Euclid satellite in conjunction with CMB  data \cite{DEC-1}. Hence next generation data have sensitivity to discrimination between early dynamical dark energy and semi- $\Lambda$CDM models.

\newpage
\acknowledgments
%We wish to appreciation to Sadegh Movahed  for discussion and communication.
 We gratefully thanks Marzieh Farhang for helpful comment and discussion in this work.  The authors would like to Lucca Amendoal and Ruth Durrer for helpful comment through Meeting in Modified Gravity in Tehran on 2016. Also the authors are appreciate from Alireza Vafaee Sadre for helpful comment in programming.


\begin{thebibliography}{99}

\bibitem{Riess:1998cb}
  A.~G.~Riess {\it et al.} [Supernova Search Team Collaboration],
 \textit{Observational evidence from supernovae for an accelerating universe and a cosmological constant},
  Astron.\ J.\  \textbf{ 11} (1998)  1009
 [astro-ph/9805201].


\bibitem{Perlmutter:1998np}
  S.~Perlmutter {\it et al.} [Supernova Cosmology Project Collaboration],
\textit{Measurements of Omega and Lambda from 42 high redshift supernovae},'
  Astrophys.\ J.\  {\bf 517} (1999) 565
  [astro-ph/9812133].
  %%CITATION = doi:10.1086/307221;%%
  %10585 citations counted in INSPIRE as of 03 Jan 2018


 \bibitem{BAO-H-1}
C.~Cheng and Q.~G.~Huang,
\textit{An accurate determination of the Hubble constant from Baryon Acoustic Oscillation datasets},
  Sci.\ China Phys.\ Mech.\ Astron.\  {\bf 58} (2015) no.9,  599801
 % doi:10.1007/s11433-015-5684-5
  [astro-ph/1409.6119].
  %%CITATION = doi:10.1007/s11433-015-5684-5;%%



\bibitem{BAO-H-2}
 Beutler, Florian, et al. \textit{The 6dF Galaxy Survey: baryon acoustic oscillations and the local Hubble constant}.
 Mon.\ Not.\ Roy.\ Astron.\ Soc.\  {\bf 416.4 } (2011) 3017.


 \bibitem{CCP}
  J.~Martin,
\textit{Everything You Always Wanted To Know About The Cosmological Constant Problem (But Were Afraid To Ask)}, Comptes Rendus Physique {\bf 13} (2012) 566
 % doi:10.1016/j.crhy.2012.04.008
  [astro-ph/1205.3365].


 \bibitem{Kunz:2012aw}
  M.~Kunz,
\textit{The phenomenological approach to modeling the dark energy},
  Comptes Rendus Physique {\bf 13} (2012) 539
  %doi:10.1016/j.crhy.2012.04.007
[astro-ph/1204.5482].

\bibitem{moore99}
 B.~Moore, S.~Ghigna, F.~Governato, G.~Lake, T.~R.~Quinn, J.~Stadel and P.~Tozzi,
\textit{Dark matter substructure within galactic halos},
  Astrophys.\ J.\  {\bf 524} (1999) L19
 % doi:10.1086/312287
  [astro-ph/9907411].

\bibitem{Klypin99}
 A.~A.~Klypin, A.~V.~Kravtsov, O.~Valenzuela and F.~Prada,
\textit{Where are the missing Galactic satellites?},
  Astrophys.\ J.\  {\bf 522} (1999) 82
 % doi:10.1086/307643
  [astro-ph/9901240].

 \bibitem{Navarro96}
J.~F.~Navarro, C.~S.~Frenk and S.~D.~M.~White,
\textit{The Structure of cold dark matter halos},
  Astrophys.\ J.\  {\bf 462} (1996) 563
%  doi:10.1086/177173
  [astro-ph/9508025].

\bibitem{Blok10} W. J. G. de Blok,\textit{The Core-Cusp Problem}, Adv. Astron. 2010, (2010) 789293.



\bibitem{Donato09} F. Donato et al.,\textit{ A constant dark matter halo surface density in galaxies},
 Mon.\ Not.\ Roy.\ Astron.\ Soc.\  {\bf 397 } (2009) 1169.


\bibitem{Adecluster16}P. A. R. Ade et al. [Planck Collaboration], \textit{Planck 2015 results. XXIV. Cosmology from Sunyaev-Zeldovich cluster counts}, Astron.\ Astrophys.\  {\bf 594} (2016) A24.


  \bibitem{Kamenshchik:2001cp}
  A.~Y.~Kamenshchik, U.~Moschella and V.~Pasquier,
 \textit{ An Alternative to quintessence,}
  Phys.\ Lett.\ B {\bf 511} (2001)  265
  %doi:10.1016/S0370-2693(01)00571-8
  [gr-qc/0103004].
  %%CITATION = doi:10.1016/S0370-2693(01)00571-8;%%
  %1398 citations counted in INSPIRE as of 05 Oct 2016

\bibitem{ArmendarizPicon:2000ah}
  C.~Armendariz-Picon, V.~F.~Mukhanov and P.~J.~Steinhardt,
\textit{ Essentials of k essence},
  Phys.\ Rev.\ D {\bf 63} (2001) 103510
 % doi:10.1103/PhysRevD.63.103510
[astro-ph/0006373].
  %%CITATION = doi:10.1103/PhysRevD.63.103510;%%
  %1084 citations counted in INSPIRE as of 05 Oct 2016

 \bibitem{Caldwell:1999ew}
  R.~R.~Caldwell,
\textit{A Phantom menace},
  Phys.\ Lett.\ B {\bf 545} (2002) 23
 % doi:10.1016/S0370-2693(02)02589-3
[astro-ph/9908168].
  %%CITATION = doi:10.1016/S0370-2693(02)02589-3;%%
  %2033 citations counted in INSPIRE as of 05 Oct 2016


\bibitem{Amendola:1999er}
  L.~Amendola,
\textit{Coupled quintessence},
  Phys.\ Rev.\ D {\bf 62} (2000) 043511
  %doi:10.1103/PhysRevD.62.043511
[astro-ph/9908023].
  %%CITATION = doi:10.1103/PhysRevD.62.043511;%%
  %1036 citations counted in INSPIRE as of 05 Oct 2016

 \bibitem{Carroll:2003wy}
  S.~M.~Carroll, V.~Duvvuri, M.~Trodden and M.~S.~Turner,
\textit{Is cosmic speed - up due to new gravitational physics?}
  Phys.\ Rev.\ D {\bf 70} (2004) 043528
 % doi:10.1103/PhysRevD.70.043528
[astro-ph/0306438].
  %%CITATION = doi:10.1103/PhysRevD.70.043528;%%
  %1501 citations counted in INSPIRE as of 09 Oct 2016



\bibitem{Nojiri:2003ft}
  S.~Nojiri and S.~D.~Odintsov,
\textit{Modified gravity with negative and positive powers of the curvature: Unification of the inflation and of the cosmic acceleration
}  Phys.\ Rev.\ D {\bf 68} (2003) 123512
  %doi:10.1103/PhysRevD.68.123512
  [hep-th/0307288].
  %%CITATION = doi:10.1103/PhysRevD.68.123512;%%
  %1152 citations counted in INSPIRE as of 09 Oct 2016

 \bibitem{Amendola:1999qq}
  L.~Amendola,
 \textit{Scaling solutions in general nonminimal coupling theories}
  Phys.\ Rev.\ D {\bf 60} (1999) 043501
 % doi:10.1103/PhysRevD.60.043501
 [astro-ph/9904120].
  %%CITATION = doi:10.1103/PhysRevD.60.043501;%%
  %423 citations counted in INSPIRE as of 09 Oct 2016

  \bibitem{Uzan:1999ch}
  J.~P.~Uzan,
\textit{Cosmological scaling solutions of nonminimally coupled scalar fields}
  Phys.\ Rev.\ D {\bf 59} (1999) 123510
  %doi:10.1103/PhysRevD.59.123510
  [gr-qc/9903004].
  %%CITATION = doi:10.1103/PhysRevD.59.123510;%%
  %358 citations counted in INSPIRE as of 09 Oct 2016


%\cite{Copeland:2006wr}
\bibitem{Copeland:2006wr}
  E.~J.~Copeland, M.~Sami and S.~Tsujikawa,
\textit{Dynamics of dark energy},
  Int.\ J.\ Mod.\ Phys.\ D {\bf 15} (2006) 1753
 [hep-th/0603057].
  %%CITATION = HEP-TH/0603057;%%
  %2692 citations counted in INSPIRE as of 12 Oct 2015
%\url{http://arxiv.org/abs/hep-th/0603057}




%\cite{Nojiri:2010wj}
\bibitem{Nojiri:2010wj}
 S.~Nojiri and S.~D.~Odintsov,
\textit{Unified cosmic history in modified gravity: from F(R) theory to Lorentz non-invariant models},
  Phys.\ Rept.\  {\bf 505}, 59 (2011)
  %doi:10.1016/j.physrep.2011.04.001
 [gr-qc/1011.0544].
  %%CITATION = doi:10.1016/j.physrep.2011.04.001;%%
  %1268 citations counted in INSPIRE as of 24 Nov 2016



%\cite{Amendola:2012ys}
\bibitem{Amendola:2012ys}
  L.~Amendola {\it et al.} [Euclid Theory Working Group Collaboration],
\textit{Cosmology and fundamental physics with the Euclid satellite},
  Living Rev.\ Rel.\  {\bf 16} (2013) 6
%  doi:10.12942/lrr-2013-6
[astro-ph/1206.1225].
  %%CITATION = doi:10.12942/lrr-2013-6;%%
  %297 citations counted in INSPIRE as of 01 Jun 2016
 % \url{http://dx.doi.org/10.12942/lrr-2013-6}



%\cite{Nojiri:2006ri}
%\bibitem{Nojiri:2006ri}
 % S.~Nojiri and S.~D.~Odintsov,
  %``Introduction to modified gravity and gravitational alternative for dark energy,''
%  eConf C {\bf 0602061}, 06 (2006)
%  [Int.\ J.\ Geom.\ Meth.\ Mod.\ Phys.\  {\bf 4}, 115 (2007)]
 % doi:10.1142/S0219887807001928
 % [hep-th/0601213].
  %%CITATION = doi:10.1142/S0219887807001928;%%
  %1562 citations counted in INSPIRE as of 24 Nov 2016


\bibitem{Bull:2015stt}
  P.~Bull {\it et al.},
\textit{Beyond $\Lambda$CDM: Problems, solutions, and the road ahead},
  Phys.\ Dark Univ.\  {\bf 12} (2016) 56
 % doi:10.1016/j.dark.2016.02.001
  [astro-ph/1512.05356].
  %%CITATION = doi:10.1016/j.dark.2016.02.001;%%
  %52 citations counted in INSPIRE as of 24 Nov 2016
%%%%%%%%%%%%%%%%%%%%%%%%%%%%%%%%%%%%%%%%%

\bibitem{Horndeski:1974wa}
G.W. Horndeski,\textit{Second-order scalar-tensor field equations in a
  four-dimensional space},''  Int. J. Theor. Phys. \textbf{10} (1974) 363.


\bibitem{Konnig:2016idp}
 F.~Könnig, H.~Nersisyan, Y.~Akrami, L.~Amendola and M.~Zumalacárregui,
\textit{A spectre is haunting the cosmos: Quantum stability of massive gravity with ghosts},
  JHEP {\bf 1611} (2016) 118
  %doi:10.1007/JHEP11(2016)118
  [gr-qc/1605.08757].
  %%CITATION = doi:10.1007/JHEP11(2016)118;%%
  %6 citations counted in INSPIRE as of 19 Jan 2018


\bibitem{Bento:2002ps}
M.~C.~Bento, O.~Bertolami and A.~A.~Sen,
\textit{Generalized Chaplygin gas, accelerated expansion and dark energy matter unification},
  Phys.\ Rev.\ D {\bf 66} (2002) 043507
  %doi:10.1103/PhysRevD.66.043507
  [gr-qc/0202064].
  %%CITATION = doi:10.1103/PhysRevD.66.043507;%%
  %1217 citations counted in INSPIRE as of 19 Jan 2018

\bibitem{Mostaghel:2016lcd}
  B.~Mostaghel, H.~Moshafi and S.~M.~S.~Movahed,
\textit{Non-minimal Derivative Coupling Scalar Field and Bulk Viscous Dark Energy},
  Eur.\ Phys.\ J.\ C {\bf 77} (2017)541
  %doi:10.1140/epjc/s10052-017-5085-1
  [astro-ph/1611.08196].
  %%CITATION = doi:10.1140/epjc/s10052-017-5085-1;%%


\bibitem{nima16}
  N.~Khosravi,
\textit{Ensemble Average Theory of Gravity},
  Phys.\ Rev.\ D {\bf 94} (2016) 124035
 % doi:10.1103/PhysRevD.94.124035
  [gr-qc/1606.01887].
  %%CITATION = doi:10.1103/PhysRevD.94.124035;%%
  %1 citations counted in INSPIRE as of 31 Jul 2017


\bibitem{nima17} Khosravi, Nima.  \textit{Uber-Gravity and the Cosmological Constant Problem}, [arXiv:1703.02052].



\bibitem{Huterer:1998qv}
  D.~Huterer and M.~S.~Turner,
\textit{Prospects for probing the dark energy via supernova distance measurements}
  Phys.\ Rev.\ D {\bf 60} (1999) 081301
 % doi:10.1103/PhysRevD.60.081301
 [astro-ph/9808133].
  %%CITATION = doi:10.1103/PhysRevD.60.081301;%%
  %258 citations counted in INSPIRE as of 09 Oct 2016

\bibitem{Sahlen:2005zw}
  M.~Sahlen, A.~R.~Liddle and D.~Parkinson,
\textit{Direct reconstruction of the quintessence potential}
  Phys.\ Rev.\ D {\bf 72} (2005) 083511
 % doi:10.1103/PhysRevD.72.083511
 [astro-ph/0506696].
  %%CITATION = doi:10.1103/PhysRevD.72.083511;%%
  %65 citations counted in INSPIRE as of 10 Oct 201

\bibitem{Nair:2013sna}
  R.~Nair, S.~Jhingan and D.~Jain,
\textit{Exploring scalar field dynamics with Gaussian processes}
  JCAP {\bf 1401} (2014) 005.
  %doi:10.1088/1475-7516/2014/01/005
 [astro-ph/1306.0606].
  %%CITATION = doi:10.1088/1475-7516/2014/01/005;%%
  %10 citations counted in INSPIRE as of 10 Oct 2016

\bibitem{Holsclaw}
Holsclaw, Tracy, et al. \textit{Nonparametric reconstruction of the dark energy equation of state}, Physical Review D \textbf{82.10} (2010) 103502
[astro-ph/1009.5443 ].

\bibitem{DEC-4}
  M.~Takada,
 \textit{Can A Galaxy Redshift Survey Measure Dark Energy Clustering},
  Phys.\ Rev.\ D {\bf 74} (2006) 043505
  %doi:10.1103/PhysRevD.74.043505
  [astro-ph/0606533].
  %%CITATION = doi:10.1103/PhysRevD.74.043505;%%
  %40 citations counted in INSPIRE as of 11 Nov 2016

  \bibitem{DEC-9}
S.~Dutta and I.~Maor,
\textit{Voids of dark energy},
  Phys.\ Rev.\ D {\bf 75} (2007) 063507
  %doi:10.1103/PhysRevD.75.063507
[gr-qc/0612027].
  %%CITATION = doi:10.1103/PhysRevD.75.063507;%%
  %41 citations counted in INSPIRE as of 08 Dec 2016



\bibitem{DEC-5}
Q. Wang,  Z. and Fan, \textit{Simulation studies of dark energy clustering induced by the formation of dark matter halos},  Phys.\ Rev.\ D {\bf 85} (2012) 023002.




\bibitem{31DEC-5}
Q. Wang,  Z. and Fan, \textit{Dynamical evolution of quintessence dark energy in collapsing dark matter halos},Phys.\ Rev.\ D {\bf 79} (2009) 123012 [astro-ph/0906.3349].



\bibitem{DEC-1}
  S.~A.~Appleby, E.~V.~Linder and J.~Weller,
\textit{Cluster Probes of Dark Energy Clustering},
  Phys.\ Rev.\ D {\bf 88} (2013) 043526
  %%doi:10.1103/PhysRevD.88.043526
 [astro-ph/1305.6982].
  %%CITATION = doi:10.1103/PhysRevD.88.043526;%%
  %7 citations counted in INSPIRE as of 26 Oct 2016


%\bibitem{PlanckDE}
 %P.~A.~R.~Ade {\it et al.} [Planck Collaboration],
%\textit{Planck 2015 results. XIV. Dark energy and modified gravity}
 % &Astron.\ Astrophys.\  {\bf 594} (2016) A14
 % doi:10.1051/0004-6361/201525814
 % [astro-ph/1502.01590].
  %%CITATION = doi:10.1051/0004-6361/201525814;%%
  %215 citations counted in INSPIRE as of 08 Dec 2016

  \bibitem{jain03}
  B.~Jain and A.~Taylor,
\textit{Cross-correlation tomography: measuring dark energy evolution with weak lensing},
  Phys.\ Rev.\ Lett.\  {\bf 91} (2003) 141302
 % doi:10.1103/PhysRevLett.91.141302
  [astro-ph/0306046].
  %%CITATION = doi:10.1103/PhysRevLett.91.141302;%%
  %294 citations counted in INSPIRE as of 19 Jan 2018

\bibitem{taylor07}A.N. Taylor, T.D. Kitching, D.J. Bacon, A.F. Heavens, \textit{Probing dark energy with the shear-ratio geometric test},
  Mon.\ Not.\ Roy.\ Astron.\ Soc.\  {\bf 374 } (2007) 1377.


\bibitem{Kitching08} T.D. Kitching, A.N. Taylor, A.F. Heavens, \textit{Systematic effects on dark energy from 3D weak shear},
 Mon.\ Not.\ Roy.\ Astron.\ Soc.\  {\bf  389 } (2008) 173.




\bibitem{Sahni:2002fz}
  V.~Sahni, T.~D.~Saini, A.~A.~Starobinsky and U.~Alam,
\textit{Statefinder: A New geometrical diagnostic of dark energy}
  JETP Lett. \textbf{77} (2003) 201
  % [Pisma Zh.\ Eksp.\ Teor.\ Fiz.\  {\bf 77} (2003) 249]
  %doi:10.1134/1.1574831
  [astro-ph/0201498].
  %%CITATION = doi:10.1134/1.1574831;%%
  %504 citations counted in INSPIRE as of 07 Oct 2016

 \bibitem{Arabsalmani:2011fz}
  M.~Arabsalmani and V.~Sahni,
\textit{The Statefinder hierarchy: An extended null diagnostic for concordance cosmology}
  Phys.\ Rev.\ D {\bf 83} (2011) 043501
  %doi:10.1103/PhysRevD.83.043501
  [astro-ph/1101.3436].
  %%CITATION = doi:10.1103/PhysRevD.83.043501;%%
  %34 citations counted in INSPIRE as of 07 Oct 2016

\bibitem{FutureGrowsFuvnction}
A. F. Heavens, T. D. Kitching, and L. Verde,\textit{ On model selection forecasting, Dark Energy and modified
gravity}, Mon.Not.Roy.Astron.Soc.\textbf{380} (2007) 1029
[astro-ph/0703191v2].

\bibitem{DES}
 T.~Abbott {\it et al.} [DES Collaboration], \textit{The dark energy survey},
  [astro-ph/0510346].
  %%CITATION = ASTRO-PH/0510346;%%
  %428 citations counted in INSPIRE as of 19 Jan 2018
  %%%%%%%%%%%%%%%%%%%%%%%%

\bibitem{Euclid}
R. Laureijs et al., \textit{Euclid Definition Study Report}, [arXiv:1606.00180] .

\bibitem{SKA}
R. DeBoerr,  R. David, et al. \textit{Australian SKA pathfinder: A high-dynamic range wide-field of view survey telescope}. Proceedings of the IEEE \textbf{97.8} (2009) 1507.

\bibitem{PRISM}
P.~Andre {\it et al.} [PRISM Collaboration],
\textit{PRISM (Polarized Radiation Imaging and Spectroscopy Mission): A White Paper on the Ultimate Polarimetric Spectro-Imaging of the Microwave and Far-Infrared Sky},
  [astro-ph/1306.2259].
  %%CITATION = ARXIV:1306.2259;%%
  %116 citations counted in INSPIRE as of 20 Jan 2018


\bibitem{CoRE}
 C. Armitage-Caplan, et al. \textit{COrE (Cosmic Origins Explorer) A White Paper}, [ arXiv:1102.2181] .

 \bibitem{SF-C-35}
T. Denkiewicz, \emph{Dynamical dark energy models with singularities in the view of the forthcoming results of the growth observations} [arXiv:1511.04708]

\bibitem{MB-5}
 R.~R.~Caldwell and E.~V.~Linder,
\textit{The Limits of quintessence},
  Phys.\ Rev.\ Lett.\  {\bf 95} (2005) 141301
  %doi:10.1103/PhysRevLett.95.141301
  [astro-ph/0505494].
  %%CITATION = doi:10.1103/PhysRevLett.95.141301;%%
  %350 citations counted in INSPIRE as of 20 Jan 2018


\bibitem{47}
Brax, Philippe, and Patrick Valageas. \textit{Goldstone models of modified gravity}, Phys.\ Rev.\ D\textbf{ 95.4} (2017) 043515.

\bibitem{48}
M.~Honda, N.~Iizuka, A.~Tanaka and S.~Terashima,
\textit{Exact Path Integral for 3D Higher Spin Gravity},
  Phys.\ Rev.\ D {\bf 95} (2017) 046016
 % doi:10.1103/PhysRevD.95.046016
  [hep-th/1511.07546].
  %%CITATION = doi:10.1103/PhysRevD.95.046016;%%
  %3 citations counted in INSPIRE as of 20 Jan 2018


\bibitem{Sf-4}
T.~Koivisto and D.~F.~Mota,
Dark energy anisotropic stress and large scale structure formation,
  Phys.\ Rev.\ D {\bf 73} (2006) 083502
  %doi:10.1103/PhysRevD.73.083502
  [astro-ph/0512135].
  %%CITATION = doi:10.1103/PhysRevD.73.083502;%%
  %176 citations counted in INSPIRE as of 20 Jan 2018

 \bibitem{MB-3}
  B.~A.~Bassett, P.~S.~Corasaniti and M.~Kunz,
\textit{The Essence of quintessence and the cost of compression},
  Astrophys.\ J.\  {\bf 617} (2004) L1
  %doi:10.1086/427023
  [astro-ph/0407364].
  %%CITATION = doi:10.1086/427023;%%
  %167 citations counted in INSPIRE as of 20 Jan 2018


%\cite{Chevallier:2000qy}
\bibitem{Chevallier:2000qy}
  M.~Chevallier and D.~Polarski,
\textit{Accelerating universes with scaling dark matter},
  Int.\ J.\ Mod.\ Phys.\ D {\bf 10} (2001) 213
  %doi:10.1142/S0218271801000822
  [gr-qc/0009008].
  %%CITATION = doi:10.1142/S0218271801000822;%%
  %1094 citations counted in INSPIRE as of 29 Apr 2017

  \bibitem{M-1}
 E.~V.~Linder,
\textit{Exploring the expansion history of the universe},
  Phys.\ Rev.\ Lett.\  {\bf 90} (2003) 091301
  %doi:10.1103/PhysRevLett.90.091301
  [astro-ph/0208512].
  %%CITATION = doi:10.1103/PhysRevLett.90.091301;%%
  %1104 citations counted in INSPIRE as of 20 Jan 2018


\bibitem{MB-2}
J.~Z.~Ma and X.~Zhang,
\textit{Probing the dynamics of dark energy with novel parametrizations},
  Phys.\ Lett.\ B {\bf 699} (2011) 233
 % doi:10.1016/j.physletb.2011.04.013
  [astro-ph/1102.2671].
  %%CITATION = doi:10.1016/j.physletb.2011.04.013;%%
  %57 citations counted in INSPIRE as of 20 Jan 2018

%\cite{Salzano:2012zp}
\bibitem{Salzano:2012zp}
  V.~Salzano, Y.~Wang, I.~Sendra and R.~Lazkoz,
\textit{Linear dark energy equation of state revealed by supernovae},
  Mod.\ Phys.\ Lett.\ A {\bf 29} (2014) 1450008
  %doi:10.1142/S0217732314500084
[astro-ph/1211.1012].
  %%CITATION = doi:10.1142/S0217732314500084;%%
  %5 citations counted in INSPIRE as of 29 Apr 2017

\bibitem{M-5}
C.~J.~Feng, X.~Y.~Shen, P.~Li and X.~Z.~Li,
\textit{A New Class of Parametrization for Dark Energy without Divergence},
  JCAP {\bf 1209} (2012) 023
 % doi:10.1088/1475-7516/2012/09/023
  [astro-ph/1206.0063].
  %%CITATION = doi:10.1088/1475-7516/2012/09/023;%%
  %19 citations counted in INSPIRE as of 20 Jan 2018

  \bibitem{M-3}
S.~Rahvar and M.~S.~Movahed,
\textit{ Power-law Parameterized Quintessence Model},
  Phys.\ Rev.\ D {\bf 75} (2007) 023512
%  doi:10.1103/PhysRevD.75.023512
[astro-ph/0604206].

\bibitem{julian67} W.H. Julian, \textit{On the effect of interstellar material on stellar non- circular velocities in disk galaxies},   Astrophys.\ J.\  {\bf 148} (1967) 175.


 \bibitem{PER-1}
  C.~P.~Ma and E.~Bertschinger,
\textit{Cosmological perturbation theory in the synchronous and conformal Newtonian gauges},
  Astrophys.\ J.\  {\bf 455} (1995) 7
  %doi:10.1086/176550
  [astro-ph/9506072].
  %%CITATION = doi:10.1086/176550;%%
  %1050 citations counted in INSPIRE as of 29 Oct 2016

  \bibitem{SF-9}
Sapone, Domenico, and Martin Kunz. \textit{Fingerprinting dark energy},  Phys.\ Rev.\ D \textbf{80.8} (2009) 083519 [astro-ph/0909.0007].

  \bibitem{POS-M-2}
A. Meiksin and M. White\textit{The growth of correlations in the matter power spectrum}, Mon.\ Not.\ Roy.\ Astron.\ Soc.\  {\bf 308 } (1999)1179.


 \bibitem{Pouri:2014nta}
  A.~Pouri, S.~Basilakos and M.~Plionis,
\textit{Precision growth index using the clustering of cosmic structures and growth data},
  JCAP {\bf 1408} (2014) 042
  %doi:10.1088/1475-7516/2014/08/042
 [astro-ph/1402.0964].
  %%CITATION = doi:10.1088/1475-7516/2014/08/042;%%
  %14 citations counted in INSPIRE as of 26 Oct 2016

\bibitem{ma99}
C.~P.~Ma, R.~R.~Caldwell, P.~Bode and L.~M.~Wang,
\textit{The mass power spectrum in quintessence cosmological models},
  Astrophys.\ J.\  {\bf 521} (1999) L1
 % doi:10.1086/312183
  [astro-ph/9906174].
  %%CITATION = doi:10.1086/312183;%%
  %150 citations counted in INSPIRE as of 20 Jan 2018

\bibitem{dodelson} S. Dodelson, \textit{Modern Cosmology }, Academic Press, New York (2008).

\bibitem{bardeen86} J.M. Bardeen, J.R. Bond, N. Kaiser, and A.S. Szalay,\textit{The statistics of peaks of Gaussian random fields},
Astrophys.\ J.\  {\bf 304} (1986)15 .

 %\cite{Ade:2015xua}
\bibitem{Ade:2015xua}
  P.~A.~R.~Ade {\it et al.} [Planck Collaboration],
 \textit{Planck 2015 results. XIII. Cosmological parameters},
  Astron.\ Astrophys.\  {\bf 594} (2016) A13
 % doi:10.1051/0004-6361/201525830
  [astro-ph/1502.01589].
  %%CITATION = doi:10.1051/0004-6361/201525830;%%
  %4006 citations counted in INSPIRE as of 14 Sep 2017

\bibitem{song09}
Y.~S.~Song and W.~J.~Percival,
Reconstructing the history of structure formation using Redshift Distortions,
  JCAP {\bf 0910} (2009) 004
  %doi:10.1088/1475-7516/2009/10/004
  [astro-ph0807.0810].
  %%CITATION = doi:10.1088/1475-7516/2009/10/004;%%
  %185 citations counted in INSPIRE as of 31 Jul 2017

 \bibitem{Kaiser:1987qv}
 Kaiser, Nick. \textit{Clustering in real space and in redshift space},  Mon.\ Not.\ Roy.\ Astron.\ Soc.\  {\bf 227.1 } (1987) 1.


\bibitem{Nesseris:2017vor}
S.~Nesseris, G.~Pantazis and L.~Perivolaropoulos,
\textit{Tension and constraints on modified gravity parametrizations of $G_{\textrm{eff}}(z)$ from growth rate and Planck data},
  Phys.\ Rev.\ D {\bf 96} (2017) 023542
  %doi:10.1103/PhysRevD.96.023542
  [astro-ph/1703.10538].
  %%CITATION = doi:10.1103/PhysRevD.96.023542;%%
  %12 citations counted in INSPIRE as of 20 Jan 2018


  \bibitem{ISW-3}
Weller, Jochen, and A. M. Lewis. \textit{Large-scale cosmic microwave background anisotropies and dark energy},
 Mon.\ Not.\ Roy.\ Astron.\ Soc.\  {\bf 346.3 } (2003) 987.


  \bibitem{sw67}
R.~K.~Sachs and A.~M.~Wolfe,
\textit{Perturbations of a cosmological model and angular variations of the microwave background},
  Astrophys.\ J.\  {\bf 147} (1967) 73.
 % [Gen.\ Rel.\ Grav.\  {\bf 39}, 1929 (2007)].
  %doi:10.1007/s10714-007-0448-9
  %%CITATION = doi:10.1007/s10714-007-0448-9;%%
  %1316 citations counted in INSPIRE as of 31 Jul 2017

  \bibitem{hu95}
 W.~Hu and N.~Sugiyama,
\textit{Anisotropies in the cosmic microwave background: An Analytic approach},
  Astrophys.\ J.\  {\bf 444} (1995) 489
  %doi:10.1086/175624
  [astro-ph/9407093].
  %%CITATION = doi:10.1086/175624;%%
  %347 citations counted in INSPIRE as of 31 Jul 2017

 \bibitem{ISW-2}
Gold, Benjamin. \textit{Limits of dark energy measurements from correlations of CMB lensing, the integrated Sachs-Wolfe effect, and galaxy counts},  Phys.\ Rev.\ D {\bf 71.6} (2005)  063522.

\bibitem{ISW-C-3}
  G.~Olivares, F.~Atrio-Barandela and D.~Pavon,
\textit{The Integrated Sachs-Wolfe Effect in Interacting Dark Energy Models},
  Phys.\ Rev.\ D {\bf 77} (2008) 103520
  %doi:10.1103/PhysRevD.77.103520
 [astro-ph/0801.4517].
  %%CITATION = doi:10.1103/PhysRevD.77.103520;%%
  %35 citations counted in INSPIRE as of 20 Oct 2016

\bibitem{DEC-2}
  D.~F.~Mota, D.~J.~Shaw and J.~Silk,
\textit{On the Magnitude of Dark Energy Voids and Overdensities},
 Astrophys.\ J.\  {\bf 675} (2008) 29
  %doi:10.1086/524401
 [astro-ph/0709.2227].
  %%CITATION = doi:10.1086/524401;%%
  %51 citations counted in INSPIRE as of 31 Jan 2018

 \bibitem{DEC-6}
P. Roland, D. Huterer, and Eric V. Linder. \textit{Measuring the speed of dark: detecting dark energy perturbations}, Phys.\ Rev.\ D {\bf 81.10} (2010) 103513  [astro-ph/1002.1311].



%\cite{Chenxiaoji:2014mxa}
\bibitem{Chenxiaoji:2014mxa}
 C.~Ling, Q.~Wang, R.~Li, B.~Li, J.~Wang and L.~Gao,
\textit{Distinguishing general relativity and $f(R)$ gravity with the gravitational lensing Minkowski functionals},
 Phys.\ Rev.\ D {\bf 92} (2015) 064024
 % doi:10.1103/PhysRevD.92.064024
  [astro-ph/1410.2734].
  %%CITATION = doi:10.1103/PhysRevD.92.064024;%%
  %4 citations counted in INSPIRE as of 20 Nov 2017


%\cite{Shirasaki:2016twn}
\bibitem{Shirasaki:2016twn}
  M.~Shirasaki, T.~Nishimichi, B.~Li and Y.~Higuchi,
\textit{The imprint of f(R) gravity on weak gravitational lensing ? II. Information content in cosmic shear statistics},
  Mon.\ Not.\ Roy.\ Astron.\ Soc.\  {\bf 466} (2017) 2402
  %doi:10.1093/mnras/stw3254
  [astro-ph/1610.03600].
  %%CITATION = doi:10.1093/mnras/stw3254;%%
  %3 citations counted in INSPIRE as of 20 Nov 2017

%\cite{Fang:2017daj}
\bibitem{Fang:2017daj}
  W.~Fang, B.~Li and G.~B.~Zhao,
\textit{New Probe of Departures from General Relativity Using Minkowski Functionals},
  Phys.\ Rev.\ Lett.\  {\bf 118}  (2017) 181301
  %doi:10.1103/PhysRevLett.118.181301
  [astro-ph/1704.02325].
  %%CITATION = doi:10.1103/PhysRevLett.118.181301;%%
  %2 citations counted in INSPIRE as of 20 Nov 2017


%\bibitem{King:2013csa}
 % A.~L.~King, T.~M.~Davis, K.~Denney, M.~Vestergaard and D.~Watson,
%High Redshift Standard Candles: Predicted Cosmological Constraints,
 % Mon.\ Not.\ Roy.\ Astron.\ Soc.\  {\bf 441} (2014) 3454
  %doi:10.1093/mnras/stu793
 %[astro-ph/1311.2356].
  %%CITATION = doi:10.1093/mnras/stu793;%%
  %13 citations counted in INSPIRE as of 29 Apr 2017


  %\cite{Akarsu:2015wqa}
%\bibitem{Akarsu:2015wqa}
  %.~Akarsu, M.~Bouhmadi-Lpez, M.~Brilenkov, R.~Brilenkov, M.~Eingorn and A.~Zhuk,
%\textit{Are dark energy models with variable EoS parameter $w$ compatible with the late inhomogeneous Universe},
%  JCAP {\bf 1507} (2015) 038
  %doi:10.1088/1475-7516/2015/07/038
 % [gr-qc/1502.04693].
  %%CITATION = doi:10.1088/1475-7516/2015/07/038;%%
  %10 citations counted in INSPIRE as of 29 Apr 2017


%\bibitem{87}
% A.~Marcos-Caballero, E.~Martínez-González and P.~Vielva,
%\textit{ Local properties of the large-scale peaks of the CMB temperature},
%  JCAP {\bf 1705} (2017) 023
 % doi:10.1088/1475-7516/2017/05/023
%  [astro-ph/1701.08552].
  %%CITATION = doi:10.1088/1475-7516/2017/05/023;%%
  %1 citations counted in INSPIRE as of 20 Jan 2018



%\cite{Gomez-Valent:2014rxa}
%\bibitem{Gomez-Valent:2014rxa}
%  A.~Gmez-Valent, J.~Sol and S.~Basilakos,
%\textit{Dynamical vacuum energy in the expanding Universe confronted with observations: a dedicated study},
%  JCAP {\bf 1501}, 004 (2015)
 % doi:10.1088/1475-7516/2015/01/004
 %[astro-ph/1409.7048].
  %%CITATION = doi:10.1088/1475-7516/2015/01/004;%%
  %47 citations counted in INSPIRE as of 12 Jul 2017

  \bibitem{Bond and efstathiou 1987}
Bond, J. R., and G. Efstathiou. \textit{The statistics of cosmic background radiation fluctuations},
 Mon.\ Not.\ Roy.\ Astron.\ Soc.\  {\bf 226 } (1987) 655.


\bibitem{ravi99}
  A.~F.~Heavens and R.~K.~Sheth,
\textit{The correlation of peaks in the microwave background},
  Mon.\ Not.\ Roy.\ Astron.\ Soc.\  {\bf 310}, 1062 (1999)
 % doi:10.1046/j.1365-8711.1999.03015.x
  [astro-ph/9904307].
  %%CITATION = doi:10.1046/j.1365-8711.1999.03015.x;%%
  %36 citations counted in INSPIRE as of 03 Nov 2016

\bibitem{Hikage:2006fe}
  C.~Hikage, E.~Komatsu and T.~Matsubara,
\textit{Primordial Non-Gaussianity and Analytical Formula for Minkowski Functionals of the Cosmic Microwave Background and Large-scale Structure},
  Astrophys.\ J.\  {\bf 653} (2006)  11
  %doi:10.1086/508653
  [astro-ph/0607284].
  %%CITATION = doi:10.1086/508653;%%
  %86 citations counted in INSPIRE as of 13 May 2017


\bibitem{ries04}
  A.~G.~Riess {\it et al.} [Supernova Search Team],
\textit{Type Ia supernova discoveries at z > 1 from the Hubble Space Telescope: Evidence for past deceleration and constraints on dark energy evolution},
  Astrophys.\ J.\  {\bf 607} (2004) 665
  %doi:10.1086/383612
  [astro-ph/0402512].
  %%CITATION = doi:10.1086/383612;%%
  %3007 citations counted in INSPIRE as of 31 Jul 2017

\bibitem{ries07}
  A.~G.~Riess {\it et al.},
\textit{New Hubble Space Telescope Discoveries of Type Ia Supernovae at z>=1: Narrowing Constraints on the Early Behavior of Dark Energy},
  Astrophys.\ J.\  {\bf 659} (2007) 98
  %doi:10.1086/510378
 [astro-ph/0611572].
  %%CITATION = doi:10.1086/510378;%%
  %1239 citations counted in INSPIRE as of 31 Jul 2017

\bibitem{ast06}
  P.~Astier {\it et al.} [SNLS Collaboration],
\textit{The Supernova legacy survey: Measurement of omega(m), omega(lambda) and W from the first year data set},
  Astron.\ Astrophys.\  {\bf 447} (2006) 31
 % doi:10.1051/0004-6361:20054185
  [astro-ph/0510447].
  %%CITATION = doi:10.1051/0004-6361:20054185;%%
  %1881 citations counted in INSPIRE as of 31 Jul 2017

\bibitem{bau08}
 S.~Baumont {\it et al.} [SNLS Collaboration],
PHotometry Assisted Spectral Extraction (PHASE) and identification of SNLS supernovae,
  Astron.\ Astrophys.\  {\bf 491} (2008) 567
  %doi:10.1051/0004-6361:200810210
 [astro-ph/0809.4407].
  %%CITATION = doi:10.1051/0004-6361:200810210;%%
  %12 citations counted in INSPIRE as of 31 Jul 2017

\bibitem{reg09}
 N.~Regnault {\it et al.} [SNLS Collaboration],
Photometric Calibration of the Supernova Legacy Survey Fields,
  Astron.\ Astrophys.\  {\bf 506} (2009) 999.
  %doi:10.1051/0004-6361/200912446
 [astro-ph/0908.3808].
  %%CITATION = doi:10.1051/0004-6361/200912446;%%
  %63 citations counted in INSPIRE as of 31 Jul 2017

 \bibitem{guy10}
J.~Guy {\it et al.} [SNLS Collaboration],
\textit{The Supernova Legacy Survey 3-year sample: Type Ia Supernovae photometric distances and cosmological constraints},
  Astron.\ Astrophys.\  {\bf 523} (2010) A7
  %doi:10.1051/0004-6361/201014468
  [astro-ph/1010.4743].
  %%CITATION = doi:10.1051/0004-6361/201014468;%%
  %179 citations counted in INSPIRE as of 31 Jul 2017

\bibitem{mik07}
%\bibitem{Miknaitis:2007jd}
  G.~Miknaitis {\it et al.},
\textit{The ESSENCE Supernova Survey: Survey Optimization, Observations, and Supernova Photometry},
  Astrophys.\ J.\  {\bf 666} (2007) 674
  %doi:10.1086/519986
  [astro-ph/0701043].
  %%CITATION = doi:10.1086/519986;%%
  %230 citations counted in INSPIRE as of 20 Jan 2018

\bibitem{wood07}
W.~M.~Wood-Vasey {\it et al.} [ESSENCE Collaboration],
\textit{Observational Constraints on the Nature of the Dark Energy: First Cosmological Results from the ESSENCE Supernova Survey},
  Astrophys.\ J.\  {\bf 666} (2007) 694
  %doi:10.1086/518642
 [astro-ph/0701041].
  %%CITATION = doi:10.1086/518642;%%
  %772 citations counted in INSPIRE as of 31 Jul 2017

\bibitem{cop06}
W.~M.~Wood-Vasey {\it et al.},
The nearby supernova factory,
  New Astron.\ Rev.\  {\bf 48} (2004) 637
  %doi:10.1016/j.newar.2003.12.056
  [astro-ph/0401513].
  %%CITATION = doi:10.1016/j.newar.2003.12.056;%%
  %37 citations counted in INSPIRE as of 20 Jan 2018

\bibitem{scal09}
 R.~A.~Scalzo {\it et al.},
\textit{Nearby Supernova Factory Observations of SN 2007if: First Total Mass Measurement of a Super-Chandrasekhar-Mass Progenitor,}
  Astrophys.\ J.\  {\bf 713} (2010) 1073
  %doi:10.1088/0004-637X/713/2/1073
 [arXiv:1003.2217 [astro-ph.CO]].
  %%CITATION = doi:10.1088/0004-637X/713/2/1073;%%
  %144 citations counted in INSPIRE as of 31 Jul 2017

\bibitem{fol10}
G.~Folatelli {\it et al.},
The Carnegie Supernova Project: Analysis of the First Sample of Low-Redshift Type-Ia Supernovae,
  Astron.\ J.\  {\bf 139} (2010) 120
  %doi:10.1088/0004-6256/139/1/120
[[astro-ph/0910.3317].
  %%CITATION = doi:10.1088/0004-6256/139/1/120;%%
  %133 citations counted in INSPIRE as of 31 Jul 2017

\bibitem{fol101}
 C.~Contreras {\it et al.},
\textit{The Carnegie Supernova Project: First Photometry Data Release of Low-Redshift Type Ia Supernovae},
  Astron.\ J.\  {\bf 139} (2010) 519
 % doi:10.1088/0004-6256/139/2/519
[astro-ph/0910.3330].
  %%CITATION = doi:10.1088/0004-6256/139/2/519;%%
  %113 citations counted in INSPIRE as of 31 Jul 2017

\bibitem{lea10}
 J.~Leaman, W.~Li, R.~Chornock and A.~V.~Filippenko,
\textit{Nearby Supernova Rates from the Lick Observatory Supernova Search. I. The Methods and Database},
  Mon.\ Not.\ Roy.\ Astron.\ Soc.\  {\bf 412} (2011) 1419
  %doi:10.1111/j.1365-2966.2011.18158.x
  [astro-ph/1006.4611].
  %%CITATION = doi:10.1111/j.1365-2966.2011.18158.x;%%
  %81 citations counted in INSPIRE as of 31 Jul 2017

\bibitem{li10}
Li, Weidong, et al.\textit{ Nearby supernova rates from the Lick Observatory Supernova Search–III. The rate–size relation, and the rates as a function of galaxy Hubble type and colour},  Mon.\ Not.\ Roy.\ Astron.\ Soc.\  {\bf 412.3} (2011) 1473.


\bibitem{holtz08}
J.~A.~Holtzman {\it et al.} [SDSS Collaboration],
\textit{The Sloan Digital Sky Survey-II Photometry and Supernova IA Light Curves from the 2005 Data},
  Astron.\ J.\  {\bf 136} (2008) 2306
  %doi:10.1088/0004-6256/136/6/2306
 [astro-ph/0908.4277].
  %%CITATION = doi:10.1088/0004-6256/136/6/2306;%%
  %121 citations counted in INSPIRE as of 31 Jul 2017

\bibitem{kess09}
 R.~Kessler {\it et al.},
\textit{First-year Sloan Digital Sky Survey-II (SDSS-II) Supernova Results: Hubble Diagram and Cosmological Parameters},
  Astrophys.\ J.\ Suppl.\  {\bf 185} (2009) 32
  %doi:10.1088/0067-0049/185/1/32
  [astro-ph/0908.4274].
  %%CITATION = doi:10.1088/0067-0049/185/1/32;%%
  %447 citations counted in INSPIRE as of 31 Jul 2017

\bibitem{Suzuki12}
 N.~Suzuki {\it et al.},
\textit{The Hubble Space Telescope Cluster Supernova Survey: V. Improving the Dark Energy Constraints Above z>1 and Building an Early-Type-Hosted Supernova Sample},
  Astrophys.\ J.\  {\bf 746} (2012) 85
  %doi:10.1088/0004-637X/746/1/85
  [astro-ph/1105.3470].
  %%CITATION = doi:10.1088/0004-637X/746/1/85;%%
  %860 citations counted in INSPIRE as of 31 Jul 2017

%\cite{Cao:2014jza}
\bibitem{Cao:2014jza}
  S.~Cao and Z.~H.~Zhu,
\textit{Cosmic equation of state from combined angular diameter distances: Does the tension with
  luminosity distances exist},
  Phys.\ Rev.\ D {\bf 90} (2014) 083006
 % doi:10.1103/PhysRevD.90.083006
  [astro-ph/1410.6567].
  %%CITATION = doi:10.1103/PhysRevD.90.083006;%%
  %6 citations counted in INSPIRE as of 25 Jul 2016

\bibitem{Betoule14}
M.~Betoule {\it et al.} [SDSS Collaboration],
\textit{Improved cosmological constraints from a joint analysis of the SDSS-II and SNLS supernova samples},
  Astron.\ Astrophys.\  {\bf 568} (2014) A22
  %doi:10.1051/0004-6361/201423413
 [astro-ph/1401.4064].
  %%CITATION = doi:10.1051/0004-6361/201423413;%%
  %468 citations counted in INSPIRE as of 31 Jul 2017

 \bibitem{Ade:2015rim}
  P.~A.~R.~Ade {\it et al.} [Planck Collaboration],
\textit{Planck 2015 results. XIV. Dark energy and modified gravity},
  Astron.\ Astrophys.\  {\bf 594} (2016) A14
 % doi:10.1051/0004-6361/201525814
 [astro-ph/1502.01590].
  %%CITATION = doi:10.1051/0004-6361/201525814;%%
  %296 citations counted in INSPIRE as of 31 Jul 2017


\bibitem{Li:2010da}
  Z.~Li, P.~Wu and H.~Yu,
\textit{Examining the cosmic acceleration with the latest Union2 supernova data},
  Phys.\ Lett.\ B {\bf 695} (2011) 1.
  %doi:10.1016/j.physletb.2010.10.044

\bibitem{JLAdata1} \texttt{http://supernova.lbl.gov/Union/}

\bibitem{JLAdata2}\texttt{http://supernovae.in2p3.fr/sdss\char`_snls\char`_jla/ReadMe.html}

  \bibitem{Riess:2011yx}
  Riess, Adam G., et al. \textit{A 3\% solution: determination of the Hubble constant with the Hubble Space Telescope and Wide Field Camera 3}, Astrophys.\ J.\  {\bf  730.2} (2011) 119.



\bibitem{Padmanabhan:2012hf}
  N.~Padmanabhan, X.~Xu, D.~J.~Eisenstein, R.~Scalzo, A.~J.~Cuesta, K.~T.~Mehta and E.~Kazin,
\textit{A 2 per cent distance to $z$=0.35 by reconstructing baryon acoustic oscillations - I. Methods and application to the Sloan Digital Sky Survey},
  Mon.\ Not.\ Roy.\ Astron.\ Soc.\  {\bf 427} (2012) 2132.


 \bibitem{Anderson:2012sa}
  L.~Anderson {\it et al.},
\textit{The clustering of galaxies in the SDSS-III Baryon Oscillation Spectroscopic Survey: Baryon Acoustic Oscillations in the Data Release 9 Spectroscopic Galaxy Sample},
  Mon.\ Not.\ Roy.\ Astron.\ Soc.\  {\bf 427} (2013) 3435.

\bibitem{Blake:2012pj}
  C.~Blake {\it et al.},
\textit{The WiggleZ Dark Energy Survey: Joint measurements of the expansion and growth history at z < 1},
  Mon.\ Not.\ Roy.\ Astron.\ Soc.\  {\bf 425} (2012) 405.

\bibitem{Beutler:2011hx}
  F.~Beutler {\it et al.},
\textit{The 6dF Galaxy Survey: Baryon Acoustic Oscillations and the Local Hubble Constant},
  Mon.\ Not.\ Roy.\ Astron.\ Soc.\  {\bf 416} (2011) 3017.


%\bibitem{hinshaw12}  G.~Hinshaw {\it et al.} [WMAP Collaboration],
%\textit{Nine-Year Wilkinson Microwave Anisotropy Probe (WMAP) Observations: Cosmological Parameter Results},
%  Astrophys.\ J.\ Suppl.\  {\bf 208} (2013) 19     [astro-ph1212.5226].


%\cite{Hinshaw:2012aka}
  \bibitem{Hinshaw:2012aka}
  G.~Hinshaw {\it et al.} [WMAP Collaboration],
Nine-Year Wilkinson Microwave Anisotropy Probe (WMAP) Observations: Cosmological Parameter Results,
  Astrophys.\ J.\ Suppl.\  {\bf 208} (2013) 19
 % doi:10.1088/0067-0049/208/2/19
 [astro-ph/1212.5226].
  %%CITATION = doi:10.1088/0067-0049/208/2/19;%%
  %2144 citations counted in INSPIRE as of 31 Jul 2016


\bibitem{Lewis:1999bs}
 A.~Lewis, A.~Challinor and A.~Lasenby,
\textit{Efficient computation of CMB anisotropies in closed FRW models},
  Astrophys.\ J.\  {\bf 538} (2000) 473
  %doi:10.1086/309179
  [astro-ph/9911177].
  %%CITATION = doi:10.1086/309179;%%
  %2414 citations counted in INSPIRE as of 20 Jan 2018

%\cite{Geng:2017apd}
\bibitem{Geng:2017apd}
 C.~Q.~Geng, C.~C.~Lee and L.~Yin,
\textit{Constraints on running vacuum model with $H(z)$ and $f \sigma_8$},
  JCAP {\bf 1708} (2017)  032
 % doi:10.1088/1475-7516/2017/08/032
  [astro-ph/1704.02136].
  %%CITATION = doi:10.1088/1475-7516/2017/08/032;%%
  %4 citations counted in INSPIRE as of 20 Jan 2018

 \bibitem{data-1}
X.~W.~Duan, M.~Zhou and T.~J.~Zhang,
\textit{Testing consistency of general relativity with kinematic and dynamical probes},
 [astro-ph/1605.03947 ].
  %%CITATION = ARXIV:1605.03947;%%
  %4 citations counted in INSPIRE as of 20 Jan 2018


\bibitem{Lewis:2002ah}
  A.~Lewis and S.~Bridle,
\textit{Cosmological parameters from CMB and other data: A Monte Carlo approach},
  Phys.\ Rev.\ D {\bf 66} (2002) 103511
%  doi:10.1103/PhysRevD.66.103511
[astro-ph/0205436].

  \bibitem{H.Akaike:1974}
Akaike, Hirotugu. \textit{A new look at the statistical model identification}, IEEE transactions on automatic control \textbf{19.6} (1974) 716.

   \bibitem{G.schwarz:1978} G. Schwarz, \textit{Estimating the dimension of a model}. Ann.
Stat.\textbf{ 6} (1978) 461.


%\cite{Xu:2016grp}
\bibitem{Xu:2016grp}
Y.~Y.~Xu and X.~Zhang,
Comparison of dark energy models after Planck 2015,
  Eur.\ Phys.\ J.\ C {\bf 76} (2016) 588
  %doi:10.1140/epjc/s10052-016-4446-5
 [astro-ph/1607.06262].
  %%CITATION = doi:10.1140/epjc/s10052-016-4446-5;%%
  %16 citations counted in INSPIRE as of 31 Jul 2017



\end{thebibliography}
\end{document}